\documentclass[12pt]{article}
\usepackage{graphicx}
\usepackage{amsmath}
\usepackage{amsfonts}
\usepackage{amssymb}
\newtheorem{theorem}{Theorem}[section]

\newtheorem{lemma}[theorem]{Lemma}

\newtheorem{proposition}[theorem]{Proposition}

\newtheorem{remark}[theorem]{Remark}

\numberwithin{equation}{section}

\begin{document}

\title{The Taylor expansion at  past time-like infinity.}

\author{ 
Helmut Friedrich\\ 
Max-Planck-Institut f\"ur Gravitationsphysik\\
Am M\"uhlenberg 1\\
14476 Golm, Germany\\
}

\maketitle

{\footnotesize

\begin{abstract}

\noindent
We study the initial value problem for the conformal  field equations with data given on 
a cone ${\cal N}_p$ with vertex $p$  so that  in a suitable conformal extension
the point $p$ will represent past time-like infinity $i^-$,  the set  ${\cal N}_p \setminus \{p\}$
will represent past null infinity ${\cal J}^-$, and the freely prescribed (suitably smooth) data will acquire the meaning of the incoming  {\it radiation field} for the prospective vacuum space-time.
It is shown that:  (i) On some coordinate neighbourhood of $p$  there exist smooth fields which satisfy the conformal vacuum  field equations and induce the given data
at all orders at $p$. The Taylor coefficients  of these fields at $p$ are uniquely determined by the free data.
(ii)  On ${\cal N}_p$ there exists a unique set of fields which  induce the given free data and
satisfy the transport equations and the inner constraints induced on ${\cal N}_p$ by the conformal field equations. These fields and 
the fields which are obtained by restricting the functions considered in (i)  to 
${\cal N}_p$ coincide at all orders at $p$.

\end{abstract}

\newpage

\section{Introduction}

\vspace{.5cm}

A purely radiative,  asymptotically flat space-time should be generated solely by gravitational radiation coming in from past null infinity, extraneous information entering the space-time  at  past time-like infinity should be excluded. A natural problem to study is then the asymptotic characteristic initial value problem for the conformal vacuum field equations where data are prescribed on a cone ${\cal N}_p$ with vertex $p$, similar to the cone $\{x_{\mu}\,x^{\mu} = 0, \,\,x^0 \ge 0\}$ in Minkowski space with vertex at $x^{\mu} = 0$. It is to be arranged such
that the prospective vacuum solution admits  a smooth conformal extension 
in which the point $p$ acquires the meaning of past time-like infinity $i^-$ and the set $ {\cal N}_p \setminus \{p\}$, swept out by the future directed null geodesics through $p$, represents past null infinity ${\cal J}^-$.

As in any other  initial value problem for Einstein's field equations two different subproblems must be analysed here: (i) one needs to analyse which part of the initial  data can be prescribed freely and how the remaining data are determined on the initial set by the field equations, 
(ii) for suitably given data one has to show the existence of a smooth solution inducing these data on the initial set. In the situation indicated above both tasks are complicated  by the fact that the initial set ${\cal N}_p$ is a smooth hypersurface only away from the vertex $p$. The  notion of smoothness and the way data are given thus  require particular considerations. The present article will be concerned with problem (i), the second problem  will be dealt with  in a forthcoming article by Chru\'sciel and Paetz (\cite{chrusciel:paetz:2013}).

From the point of view of the physical/geometrical interpretation one would like 
to construct the space-times  from a minimal set of data on ${\cal N}_p$  which admit a physical interpretation.  There are various ways to prescribe data for Einstein's  field equations in characteristic initial value problems (cf. \cite{chrusciel:paetz:2012}), the specific choice usually depending on  technical considerations and the particular situation at hand. 
A natural datum to prescribe {\it at null infinity} is the {\it radiation field}, a
complex-valued function that encodes  information
on the two components of the conformal Weyl tensor with the slowest fall-off behaviour at past null infinity, thought to  represent  the two polarization states of the incoming gravitational radiation.

That the radiation field is convenient from the technical point of view has been shown in  the proof of 
J. K\'ann\'ar's existence results on the characteristic asymptotic initial value problem 
where data are prescribed on an incoming null hypersurface ${\cal C}$ which intersects past null infinity in a space-like slice $\Sigma = {\cal C} \cap {\cal J}^-$ and on the future 
$ {\cal J}'^-$ of that slice in past null infinity  \cite{kannar:1996}. A basic step in that proof consists in showing that given the radiation field on $ {\cal J}'^-$, the solution and its derivatives of any order can be determined on $ {\cal J}'^-$ by solving ODE's along the null generators of $ {\cal J}'^-$, where the initial
data for the integration are derived from the data prescribed on ${\cal C}$ and $\Sigma$.

In the problem to be considered here the analysis is complicated by the fact that
 the initial hypersurface tends to loop back onto itself near past time-like infinity, 
forcing  any analogue of $\Sigma$ to shrink to a point and leaving  no space for a hypersurface like ${\cal C}$.  The information for the integration of the solution along the null generators of 
${\cal N}_p$ has thus to be extracted completely  from the radiation field on ${\cal N}_p$. Together with the need for a careful discussion of smoothness requirements near the vertex $p$ this  leads to various algebraic subtleties.

A first study of  this problem was made  in \cite{friedrich:pure-rad:1986}, where it
was shown that for a suitably smooth prescribed radiation field  on  ${\cal N}_p$
and a gauge involving a null coordinate adapted to ${\cal N}_p$
the prospective solution to the conformal field equations is determined  uniquely  at all orders along the cone 
${\cal N}_p$. However, even under the most convenient assumptions such a null coordinate is singular at and near $p$.  To show that any smooth solution is {\it determined uniquely  in the future of the cone ${\cal N}_p$ by its radiation field},
there has been performed in \cite{friedrich:pure-rad:1986}  a transformation into a  gauge 
which is regular up to an order sufficient for the argument.
An {\it existence result for smooth solutions} would require, however,  a smooth gauge
and thus, due to the quasi-linearity of the equations, a transformation which enters the solutions at all orders. 

To simplify this tedious problem (in section \ref{transport-equ-and-inner-constr} it will be seen that the analysis of the transport equations on ${\cal N}_p$ requires a discussion of singular equations in any case) the analysis in the present article will be based on a smooth gauge right from the outset. After introducing and discussing the field equations and suitable gauge conditions in sections
\ref{conformal field equations}  - \ref{gauge-cond} the normal expansion at the point $p$ representing  space-like infinity and the properties of the radiation field on ${\cal N}_p$ are discussed in section \ref{null-data}. 

In section \ref{formal expansion} an argument by Penrose
(\cite{penrose:1980}, \cite{penrose:rindler:I}) is adapted to the present situation and it is  shown in Lemma \ref{unique-expansion} that the covariant derivatives at $p$  of the curvature fields, the conformal factor, and a further scalar field are  determined on a formal level uniquely at all orders by the radiation field and that the latter is not subject to any restriction. 
To relate these data to a space-time metric we consider in section \ref{structural equations} the structural equations, written as equations for the metric coefficients and  connection coefficients. It turns out that already a subset of the equations suffices to determine the formal Taylor expansions of these fields and that the expansion coefficients so obtained encode the information on the chosen gauge (Lemma \ref{x-ghat-normal}). 

By Borel's theorem there exist then smooth fields near $p$ whose Taylor expansion coefficients at $p$ are given precisely by the (symmetric parts of the) coefficients determined in the formal calculations above. Because only the symmetric parts of the covariant derivatives enter the definition of these functions and only a subset of the structural equation has been considered in the formal calculations, it remains to be shown that the functions so defined do indeed satisfy the conformal fields equations at all orders at $p$. A somewhat involved induction argument shows that this is  the 
case (Proposition \ref{main-result}).

The conformal field equations induce a set of inner equations on ${\cal N}_p$ which splits naturally into two subsets.  The equations in the first set, referred to as transport equations, determine all unknown fields entering the conformal field equations once the radiation field is given 
(Proposition \ref{transport-equ}).
The equations in the second set are inner constraints on the fields so determined. It turns out that they are satisfied by a solution to the transport equations without imposing restrictions on the prescribed radiation field.

To identify and analyse the inner equations on ${\cal N}_p$ one needs to express the equations in terms of a frame adapted to the cone, which is necessarily singular  at $p$. If the resulting equations are solved and the fields are then transformed back into the regular gauge underlying Proposition \ref{main-result} they coincide with the field discussed in that Proposition at all orders at $p$ and thus satisfy a necessary smoothness requirement.

The fields so obtained, which  constitute a complete set of initial data on ${\cal N}_p$ for the conformal field equations, can be considered as a starting point for an existence proof in the category of smooth functions.

As pointed out at various places, the analysis presented in this paper also applies to the characteristic initial value problem where data are given on a finite cone ${\cal N}_p$ which is thought of as being generated by the (future directed) null geodesics through a point $p$ which is considered as  an inner point of a smooth vacuum space-time. In fact, the analogues of the arguments used in sections 
\ref{conformal field equations} - \ref{f-and-f-derivatives} considerably simplify in that case.  In section \ref{transport-equ-and-inner-constr}, however,  we take advantage of the fact that the conformal Weyl tensor vanishes at null infinity. This allows us to obtain explicit expression for various fields. The analogue of Proposition \ref{transport-equ} has to be established in the finite problem by an abstract discussion of the transport equations, which will not be given here.

\section{The metric conformal field equations}
\label{conformal field equations}

\vspace{.5cm}

Let $g$ denote a Lorentzian metric on a four dimensional manifold and $\nabla$ a connection which is metric compatible so that  $\nabla g = 0$. In the following we shall make use  of a frame $\{e_k\}_{k = 0, \ldots, 3}$ which is orthonormal so that  $g_{ij} = g(e_i, e_j) = \eta_{ij}$. With  the directional covariant derivative operators $\nabla_i \equiv \nabla_{e_i}$ the connection coefficients 
$ \Gamma_i\,^k\,_j$ are define by the equation $\nabla_i\,e_j = \Gamma_i\,^k\,_j\,e_k$. The relation $\nabla g = 0$ is then equivalent to the anti-symmetry 
$\Gamma_{i\,l\,j}= - \Gamma_{i\,j\,l}$, where $\Gamma_{i\,l\,j} =  \Gamma_i\,^k\,_j\,g_{kl}$. 
All tensors (except the frame fields) will be given in the following  in terms of the frame $e_k$. 

For a vector field $Z$ the commutator of the covariant derivatives satisfies
\begin{equation}
\label{gen-comm}
(\nabla_i \nabla_j - \nabla_j \nabla_i)Z^l
= r^l\,_{kij}\,Z^k - t_i\,^k\,_j\,\nabla_kZ^l,
\end{equation}
where $t_k\,^i\,_l$ denotes the torsion tensor, given in terms of  coordinates $x^{\mu}$ and  the frame coefficients 
$e^{\mu}\,_k = \,<e_k, dx^{\mu}>$  by the relation
\begin{equation}
\label{torsion-tensor}
t_k\,^i\,_l\,e^{\mu}\,_i = e^{\mu}\,_{k,\,\nu}\,e^{\nu}\,_l - e^{\mu}\,_{l,\,\nu}\,e^{\nu}\,_k 
- (\Gamma_l\,^i\,_k - \Gamma_k\,^i\,_l)\,e^{\mu}\,_i,
\end{equation}
and $r^i\,_{jkl} $ is the curvature tensor, given by 
\begin{equation}
\label{torsion-tensor-r}
r^i\,_{jkl} \equiv
\Gamma_l\,^i\,_{j,\,\mu}\,e^{\mu}\,_k - \Gamma_k\,^i\,_{j,\,\mu}\,e^{\mu}\,_l
+ \Gamma_k\,^i\,_p\,\Gamma_l\,^p\,_j -  \Gamma_l\,^i\,_p\,\Gamma_k\,^p\,_j
\end{equation}
\[
- (\Gamma_k\,^p\,_l -  \Gamma_l\,^p\,_k
- t_k\,^p\,_l)\,\Gamma_p\,^i\,_j.
\]
The last term on the right hand side of the equation above can also be expressed in terms of the commutator of the frame fields
because $[e_k, e_l] = (\Gamma_k\,^p\,_l -  \Gamma_l\,^p\,_k - t_k\,^p\,_l)\,e_p$
by (\ref{torsion-tensor}). The metric is torsion free if and only if the torsion tensor vanishes, which is the case if and only if 
\begin{equation}
\label{torsion-free-test}
(\nabla_j\nabla_k - \nabla_k\,\nabla_j)\,f = 0,
\end{equation}
for any $C^2$-function $f$.

The torsion and the curvature tensor  satisfy in general  the Bianchi identities
\begin{equation}
\label{hat-1st-Bianchi}
\sum_{cycl(ijl)}\nabla_i\,t_j\,^k\,_l =
\sum_{cycl(ijl)}(r^k\,_{ijl} - t_i\,^m\,_j\,t\,_m\,^k\,_l),
\end{equation}
\begin{equation}
\label{hat-2nd-Bianchi}
\sum_{cycl(ijl)}\nabla_i\,r^h\,_{k j l} =
\sum_{cycl(ijl)}t_j\,^m\,_i\,r^h\,_{kml},
\end{equation}
where the sums are performed after a cyclic permutation of the indices $i,j, l$.

\vspace{.5cm}

Assume now that the metric $g$ is torsion free and related by a conformal rescaling $g = \Omega^2 \,\tilde{g}$
with a conformal factor $\Omega$ to a  `physical' metric  $\tilde{g}$ which satisfies Einstein's vacuum field equations.
These equations can then be expressed in terms of 
$g$ and $\Omega$ and derived fields  as follows. We write
\[
R_{ijkl} = C_{ijkl}
+ 2\,\{g_{i [k}\,L_{l] j} + L_{i [k}\,g_{l] j}\},
\]
where $C_{ijkl}$ is the conformal Weyl tensor and 
\[
L_{ij} 
= \frac{1}{2}\,(S_{ij} + \frac{1}{12}\,R\,g_{ij}) \quad 
\mbox{with} \quad 
S_{ij} = R_{ij} - \frac{1}{4}\,R\,g_{ij},
\]
denotes the Schouten tensor of $g$ with Ricci tensor $R_{kl}$ and Ricci scalar $R$.
In terms of the tensor fields
\[
\Omega, \quad g_{ij} = \eta_{ij}, \quad L_{ij}, \quad 
W^{i}\,_{jkl} = \Omega^{-1}\,C^{i}\,_{jkl}, \quad 
 \Pi = \frac{1}{4}\,\nabla_{i}\nabla^{i}\,\Omega + \frac{1}{24}\,R\,\Omega,
\]
the (metric) {\it conformal field equations} read
(\cite{friedrich:reg-asymp:1981a}, \cite{friedrich:asymp-sym-hyp:1981b}) 
\[
6\,\Omega\,\Pi - 3\,\nabla_{i}\Omega\,\nabla^{i}\Omega = \,0,
\]
\[
\nabla_{j}\,\nabla_{k}\,\Omega = - \Omega\,L_{jk} + \Pi\,g_{jk},
\]
\[
\nabla_{l}\, \Pi = - \nabla^{k}\Omega\,L_{k l},
\]
\[
\nabla_{i}\,L_{j k} - \nabla_{j}\,L_{i k} =
\nabla_{l}\Omega\,W^{l}\,_{k i j},
\]
\[
\nabla_{i}W^{i}\,_{jkl} = 0.
\]
These equations must be  complemented by the structural equations, namely the 
{\it torsion-free condition}
\begin{equation}
\label{torsion-free-cond}
t_k\,^i\,_l = 0,
\end{equation}
and the equation
\begin{equation}
\label{ricci-id} 
r^i\,_{jkl} = R^i\,_{jkl},
\end{equation}
which will be referred to as the {\it Ricci identity}.

\vspace{.3cm}

We note that with the choice $\Omega \equiv 1$ the conformal field equations reduce to the vacuum field equations.
The only non-trivial fields are then $e^{\mu}\,_k$, $\Gamma_i\,^j\,_k$, and  $W^{i}\,_{jkl}= C^{i}\,_{jkl}$ and the only non-trivial 
equations are the vacuum Bianchi identity $\nabla_{i}W^{i}\,_{jkl} = 0$ and the structural equations.

In the case of a more general conformal factor  the equation $6\,\Omega\, \Pi\ - 3\,\nabla_{i}\Omega\,\nabla^{i}\Omega = \,0$ will be satisfied on the connected component $C_q$ of a point $q$ if it holds at $q$ and the other equations are satisfied on $C_q$. This is a consequence of the fact that  the other equations imply the relation
\[
\nabla_{k}(6\,\Omega\, \Pi - 3\,\nabla_{j}\Omega\,\nabla^{j}\Omega) = 0.
\]
In the situations considered here,  in which either  $\Omega = 0$, $\nabla_i\Omega = 0$
or $\Omega \equiv 1$ at the point $p$,  the equation $6\,\Omega\, \Pi - 3\,\nabla_{i}\Omega\,\nabla^{i}\Omega = \,0$ need not be considered any longer.

\section{The 2-index spinor representation}
\label{spinors}

\vspace{.5cm}

The 2-index spin frame formalism is well adapted to the null geometry and  will
simplify our algebraic task considerably. It amounts essentially to  taking complex linear combinations of various expressions in terms of maps of the form
\begin{equation}
\label{alpha-map}
T_{ijk \ldots} \rightarrow T_{AA'BB'CC' \ldots} 
= T_{ijk \ldots}\,\alpha^i\,_{AA'}\,\alpha^j\,_{BB'}\,\alpha^k\,_{CC'}\,\ldots,
\end{equation}
where the $\alpha$'s denote the constant van der Waerden symbols
\[
\alpha^{i}\,_{AA'} = \frac{1}{\sqrt{2}}
\,\left(
\begin{array}{cc} 
\delta^{i}_0 + \delta^{i}_3 & \delta^{i}_1 - i\, \delta^{i}_2\\
\delta^{i}_1 + i\, \delta^{i}_2 & \delta^{i}_0  - \delta^{i}_3 
\end{array} \right), 
\]
which are hermitian matrices so that  $\overline{\alpha^k\,_{AB'}} = \alpha^k\,_{BA'}$.
Frame indices  $k,l, \ldots $ are thus replaced  by pairs of indices $AA', BB', \ldots $, where $A, B, \ldots$, $A', B', \ldots$ take values $0$ and $1$.
None of the operations applied in the following to spinor fields mix primed and unprimed indices. Therefore  we shall write $T_{ABC \ldots A'B'C' \ldots}$ instead of  $T_{AA'BB'CC' \ldots} $ if convenient.
There is an operation of complex conjugation under which unprimed indices are converted into primed indices and vice versa. Because of the hermiticity of the 
$\alpha$'s the reality of a tensor $T_{ijk \ldots} $ is then expressed by the relation
\[
\overline{T_{AA'BB'CC' \ldots} } = T_{AA'BB'CC' \ldots}.
\]
These tensor fields are considered as members of a tensor algebra which is generated by a 2-dimensional complex vector space and its primed version, both being  related to each other by an operation of complex conjugation. The members of these spaces  are called spinors. 
For more details (not in all cases employing the same curvature conventions as  used here)
we refer to \cite{penrose:rindler:I}.

The $e_k$ are also replaced by $e_{AA'} = \alpha^k\,_{AA'}\,e_k$ so that the indices $A, A'$
specify in this case the frame vector fields.
Then $e_{00'}$, $e_{11'}$ are real and $e_{01'}$, $e_{10'}$ are complex
(conjugate) null vector fields with scalar products  
\begin{equation}
\label{eAA'-scalar-prods}
g(e_{AA'},e_{BB'}) = \eta_{jk}\,\alpha^j\,_{AA'}\,\alpha^k\,_{BB'}
= \epsilon_{AB}\,\epsilon_{A'B'}, 
\end{equation}
where $\epsilon_{AC}$, $\epsilon_{A'C'}$, $\epsilon^{AC}$,
$\epsilon^{A'C'}$ denote the anti-symmetric spinor fields with 
$\epsilon_{01} = \epsilon_{0'1'} = \epsilon^{01} = \epsilon^{0'1'} =1$, so that,
assuming the summation rule for primed and unprimed indices separately,
$\epsilon_A\,^B = \epsilon_{AC}\,\epsilon^{BC}$ and 
$\epsilon_{A'}\,^{B'} = \epsilon_{A'C'}\,\epsilon^{B'C'}$ denote 
Kronecker spinors. The $\epsilon$'s are use to raise and lower indices according to the rules
\[
\kappa^A = \epsilon^{AB}\,\kappa_B, \quad  
\kappa_B =  \kappa^A\,\epsilon_{AB},
\]
and similar rules apply to  primed indices.
Upper frame indices can be converted into spinor indices by 
the  van der Waerden symbols $\alpha_{i}\,^{AA'} = 
\eta_{ij}\,\epsilon^{AB}\,\epsilon^{A'B'}\,\alpha^{j}\,_{BB'}$ 

Though it will occasionally be convenient to go back to the standard frame notation (or to employ a hybrid notation as discussed below), we shall assume most of the time the fields (except the frame and the spin frame)
to be given by their components with respect a suitably chosen spin frame field $\{\iota_{A}\}_{A = 0, 1}$ which is normalized such that
\begin{equation}
\label{spin-basis-normalization}
\epsilon(\iota_A, \iota_B) = \epsilon_{AB},
\end{equation} 
where $\epsilon$ denotes the antisymmetric form on spinor space. 
As discussed in detail in \cite{penrose:rindler:I}, the fields  
$e_{00'} = \iota_0\,\bar{\iota}_{0'}$ and 
$e_{11'} = \iota_1\,\bar{\iota}_{1'}$ correspond to real null vector fields 
while $e_{01'} = \iota_0\,\bar{\iota}_{1'}$ and 
$e_{10'} = \iota_1\,\bar{\iota}_{0'}$ correspond to complex (conjugate) null vector fields which have the  scalar products (\ref{eAA'-scalar-prods}) as a consequence of (\ref{spin-basis-normalization}).

\vspace{.3cm}

We set $\Gamma_{AA'}\,^{BB'}\,_{CC'} 
= \Gamma_i\,^j\,_k\,\alpha^i\,_{AA'}\,\alpha_j^{BB'}\,\alpha^k\,_{CC'}$. As a consequence of the anti-symmetry 
$\Gamma_{ijk} = - \Gamma_{ikj}$ these connection coefficients can be decomposed in the form
\[
\Gamma_{AA'}\,^{BB'}\,_{CC'} =
\Gamma_{AA'}\,^B\,_C\,\epsilon_{C'}\,^{B'}
+ \bar{\Gamma}_{AA'}\,^{B'}\,_{C'}\,\epsilon_C\,^B,
\] 
with spin connection coefficients  $\Gamma_{AA'}\,^B\,_C = \frac{1}{2}\,\Gamma_{AA'}\,^{BE'}\,_{CE'}$  that satisfy $\Gamma_{AA'BC} = \Gamma_{AA'(BC)}$.
Covariant derivatives of spinor fields
$\kappa^A$ resp. $\pi^{A'}$ are then defined by  
\[
\nabla_{AA'}\kappa^B = e^{\mu}_{AA'}\partial_{\mu}\,\kappa^B + \Gamma_{AA'}\,^B\,_C\,\kappa^C,
\quad
\nabla_{AA'}\pi^B = e^{\mu}_{AA'}\partial_{\mu}\,\pi^{B'} + \bar{\Gamma}_{AA'}\,^{B'}\,_{C'}\,\pi^{C'}, 
\]
and the definition of the covariant derivative is extended to arbitrary spinor fields by requiring the Leibniz rule for spinor products. For the commutators of covariant derivatives we get
\begin{equation}
\label{spin-commutator}
(\nabla_{CC'}\nabla_{DD'} - \nabla_{DD'}\nabla_{CC'}) \,\kappa^A
= R^A\,_{BCC'DD'}\,\kappa^B,
\end{equation}
and its complex conjugate,
where $R_{ABCC'DD'} = R_{(AB)CC'DD'}$ denotes  the curvature spinor. The usual curvature tensor
describing the commutator of covariant derivatives acting of vector field is then given by
\begin{equation}
\label{curvature-spin curvature}
R^{AA'}\,_{BB' CC' DD'} = 
R^i\,_{jkl} \,\alpha_i\,^{AA'}\,\alpha^j\,_{BB'}\,\alpha^k\,_{CC'}\,\alpha^l\,_{DD'}
\end{equation}
\[
= R^A\,_{BCC'DD'}\,\epsilon_{B'}\,^{A'}
+  \bar{R}^{A'}\,_{B'CC'DD'}\,\epsilon_{B}\,^{A}.
\]
The curvature spinor admits a decomposition of the form
\begin{equation}
\label{spin-curvature-decomposition}
R_{ABCC'DD'} = \Psi_{ABCD}\,\epsilon_{C'D'} + \Phi_{ABC'D'}\,\epsilon_{CD}
+ 2\,\Lambda\,\,\epsilon_{A(C}\,\epsilon_{D)B}\,\epsilon_{C'D'}.
\end{equation}
The different components are the Weyl spinor
\[
 \Psi_{ABCD} =  \Psi_{(ABCD)}
 = - C_{ijkl}\,\alpha^i\,_{AE'}\,\alpha^j\,_B\,^{E'}\,\alpha^k\,_{CF'}\,\alpha^l\,_D\,^{F'},
 \] 
which contains the information on the conformal Weyl tensor, given by
\[
C_{AA'BB'CC'DD'}= - \Psi_{ABCD}\,\epsilon_{A'B'}\,\epsilon_{C'D'}
- \bar{\Psi}_{A'B'C'D'}\,\epsilon_{AB}\,\epsilon_{CD},
\]
and the spinor
\[
\Phi_{ABA'B'} = \Phi_{(AB)(A'B')}  = \bar{\Phi}_{ABA'B'} 
= \frac{1}{2}\,(R_{jk} - \frac{1}{4}\,R\,\eta_{jk})\,\alpha^j\,_{AA'}\,\alpha^k\,_{BB'},
\]
which represents the trace free part of the Ricci tensor, and 
\[
\Lambda = \bar{\Lambda} = \frac{1}{24}\,R.
\]
It holds then
\[
L_{ABA'B'} = \Phi_{ABA'B'} + \,\Lambda\,\epsilon_{AB}\,\epsilon_{A'B'},
\]
and the rescaled conformal Weyl tensor 
$W^i\,_{jkl} = \Omega^{- 1}\,C^i\,_{jkl}$ is represented by the rescaled Weyl spinor
\[
\psi_{ABCD} = \Omega^{-1}\,\Psi_{ABCD}.
\]

\noindent
With this notation the conformal field equations read 
\[
\nabla_{AA'}\, \Pi = - \nabla^{BB'}\Omega\,(\Phi_{ABA'B'} + \,\Lambda\,\epsilon_{AB}\,\epsilon_{A'B'}),
\]
\[
\nabla_{AA'}\,\nabla_{BB'}\,\Omega = 
- \Omega\,(\Phi_{ABA'B'} + \,\Lambda\,\epsilon_{AB}\,\epsilon_{A'B'}) 
+  \Pi \,\epsilon_{AB}\,\epsilon_{A'B'},
\]
\[
\nabla_A\,^{D'}\,\Phi_{BCB'D'} + 2\,\epsilon_{A(B}\,\nabla_{C)B'}\,\Lambda
= \psi_{ABCD}\,\nabla^D\,_{B'}\Omega,
\]
\[
 \nabla^D\,_{B'}\,\psi_{ABCD} = 0.
\]
and the structural equations take the form
\[
0 =  e^{\mu}\,_{AA',\,\nu}\,e^{\nu}\,_{BB'}
 - e^{\mu}\,_{BB',\,\nu}\,e^{\nu}\,_{AA'} 
- (\Gamma_{BB'}\,^{CC'}\,_{AA'} - \Gamma_{AA'}\,^{CC'}\,_{BB'})\,e^{\mu}\,_{CC'},
\]

\[
r^A\,_{BCC'DD'} = \Omega\,\psi^A\,_{BCD}\,\epsilon_{C'D'} 
+ \Phi^A\,_{BC'D'}\,\epsilon_{CD}
+ 2\,\Lambda\,\,\epsilon^A\,_{(C}\,\epsilon_{D)B}\,\epsilon_{C'D'}.
\]
where
\begin{equation}
\label{r-Gamma-relation}
r^A\,_{BCC'DD'} = 
\end{equation}
\[
\Gamma_{DD'}\,^A\,_{B,\,\mu}\,e^{\mu}\,_{CC'}
-\Gamma_{CC'}\,^A\,_{B,\,\mu}\,e^{\mu}\,_{DD'}
+ \Gamma_{CC'}\,^A\,_F\,\Gamma_{DD'}\,^F\,_B
- \Gamma_{DD'}\,^A\,_F\,\Gamma_{CC'}\,^F\,_B
\]
\[
- ( \Gamma_{CC'}\,^{FF'}\,_{DD'} - \Gamma_{DD'}\,^{FF'}\,_{CC'})\,\Gamma_{FF'}\,^A\,_B.
\]

\vspace{.2cm}

\noindent
In the case of the vacuum field equations,  in which $\Omega \equiv 1$, the non-trivial unknowns are given by $ e^{\mu}\,_{AA'}$, $\Gamma_{AA'}\,^B\,_C$, 
$\psi_{ABCD}$ and the field equations reduce to  $\nabla^D\,_{B'}\,\psi_{ABCD} = 0$
and the structural equations.

\vspace{.5cm}

The following observations will become important later. Forget the meaning of the fields considered above and  let the spinor field $R_{ABCC'DD'}$
in (\ref{spin-curvature-decomposition}) be given by  spinor fields  
$\Psi_{ABCD}$, $ \Phi_{ABC'D'}\,\epsilon_{CD}$, and $\Lambda$ which satisfy the symmetries and reality conditions stated above. The tensor $R^{AA'}\,_{BB' CC' DD'}$
defined by (\ref{curvature-spin curvature}) then satisfies the analogue of the first  Bianchi identity  
$R^i\,_{[jkl]} = 0$ 
as a consequence of the symmetries and reality conditions.  
In fact, the anti-symmetric tensor $\epsilon_{ijkl}  = \epsilon_{[ijkl]}$ with $\epsilon_{0123} = 1$
has the spinor representation 
\[
\epsilon_{AA' BB' CC' DD'} = i\,(\epsilon_{AC}\,\epsilon_{BD}\,\epsilon_{A'D'}\,\epsilon_{B'C'}
- \epsilon_{AD}\,\epsilon_{BC}\,\epsilon_{A'C'}\,\epsilon_{B'D'}),
\] 
which implies 
\begin{equation}
\label{Bianchi-1}
R_{AA'BB'CC'DD'}\,\epsilon_{EE'}\,^{BB' CC' DD'}  
= 
2\, i\,(R_{AH}\,_{EA'}\,^H\,_{E'} -  \bar{R}_{A'H'}\,_{AE'E}\,^{H'}) = 0,
\end{equation}
because
\[
R_{AH}\,_{EA'}\,^H\,_{E'}= 
\Phi_{AEA'E'}  
- 3\,\Lambda\,\epsilon_{AE}\,\epsilon_{A'E'} = 
\bar{R}_{A'H'}\,_{AE'E}\,^{H'}.
\]
An analogue of  the second Bianchi identity $\nabla_{[m}R^{ij}\,_{kl]} = 0$ follows under suitable assumptions. It holds
\begin{equation}
\label{Bianchi-2a}
\nabla_{EE'}R_{AA'BB'CC'DD'}\,\epsilon^{EE' }\,_{FF'}\,^{CC' DD'} 
\end{equation}
\[
= 2\,i\,(\epsilon_{B'A'}\,\nabla^{CD'}R_{ABCF'FD'}
+  \epsilon_{BA}\,\nabla^{CD'}\bar{R}_{A'B'CF'FD'}),
\]
and, with $\Psi_{ABCD} = \Omega\,\psi_{ABCD}$,
\begin{equation}
\label{Bianchi-2b}
\nabla^{CD'}R_{ABCF'FD'} =  
- \Omega\,\,\nabla^{C}\,_{F'}\psi_{ABCF}
\quad \quad  \quad  \quad  \quad  \quad  
\quad \quad  \quad  \quad  \quad  \quad  
\quad \quad   
\end{equation}
\[
  \quad  \quad  \quad  
\quad \quad  \quad  \quad  \quad  \quad   + \left\{\nabla_F\,^{D'}\Phi_{ABF'D'} 
+ 2\,\epsilon_{F(A}\,\nabla_{B)F'}\Lambda - \nabla^{C}\,_{F'}\Omega\,\,\psi_{ABCF}\right\},
\]
which will vanish if the conformal field equations are satisfied. These relations are not  surprising, because the Bianchi identities have in fact been used to derive the symmetry properties of the curvature spinors and also the conformal field equations.  Later on we shall need to consider the last two relations, however, under circumstances in which it is not clear, whether the conformal field equations hold.

\vspace{.7cm}

 To shorten the following expressions it will be convenient to introduce some additional notation.
 In the case of spinor fields which carry  pairs of spinor indices like $AA'$ which correspond to a standard frame indices $j$  we shall occasionally employ a hybrid notation by using the index $j$, so that equation (\ref{r-Gamma-relation}) takes for instance  the form
\[
r^A\,_{B\,ij} = 
\]
\[
\Gamma_{j}\,^A\,_{B,\,\mu}\,e^{\mu}\,_{i}
-\Gamma_{i}\,^A\,_{B,\,\mu}\,e^{\mu}\,_{j}
+ \Gamma_{i}\,^A\,_F\,\Gamma_{j}\,^F\,_B
- \Gamma_{j}\,^A\,_F\,\Gamma_{i}\,^F\,_B
- ( \Gamma_{i}\,^{k}\,_{j} - \Gamma_{j}\,^{k}\,_{i})\,\Gamma_{k}\,^A\,_B.
\]

\vspace{.2cm}

\noindent
The symmetric part of a spinor field $S_{AB \ldots EF}$ is denoted by  $S_{(AB \ldots EF)}$.
The {\it totally symmetric part } of a spinor field
$T_{A_1 \, \ldots \, A_k\,B'_1 \ldots B_j'}$ is then given   by
$T_{(A_1 \,\ldots\,A_k)\,(B'_1 \ldots B_j')}$.
 If $T$ is a spinor field and ${\bf n} = (i_1, \ldots , i_n)$ a multi-index of order $|{\bf n}| = n$ we write
$\nabla_{\bf n}T = \nabla_{i_1} \ldots  \nabla_{i_n} T$ and 
$\nabla_{(\bf n)}T = \nabla_{(i_1} \ldots  \nabla_{i_n)} T$. If $X^i$ is a vector field we set 
$X^{\bf n} = X^{i_1} \ldots  X^{i_n}$ and write
$X^{i_1} \ldots  X^{i_n}\, \nabla_{i_1} \ldots  \nabla_{i_n} T
= X^{\bf n}\,\nabla_{\bf n}T =  X^{\bf n}\,\nabla_{(\bf n)}T $.

\section{Gauge conditions}
\label{gauge-cond}

Unless stated otherwise the connection $\nabla$ will be assumed in the following to be $g$-compatible and torsion free.
We need to restrict the gauge freedom for  the conformal factor, the frame, the coordinates.

\vspace{.4cm}

\noindent
{\it The conformal gauge near $i^-$.}

\vspace{.4cm}

 The  data for the conformal field equations are to be  prescribed on the cone 
 ${\cal N}_p = {\cal J}^- \cup \{i^-\}$.
 The vertex $p = i^-$ is to represent past time-like infinity and  ${\cal N}_p$
 is thought to be generated by the future directed null geodesics starting at $p$.
 Thus one  must assume  that 
\[ 
\Omega = 0, \quad \nabla_{AA'}\Omega = 0, \quad  \Pi  \neq 0 
\quad \mbox{at } \quad  p.
\]
The equations
$\nabla_{j}\,\nabla_{k}\,\Omega = - \Omega\,L_{jk} +  \Pi \,g_{jk}$ and 
$\nabla_l\, \Pi  = - \nabla^{k}\Omega\,L_{kl}$ suitably transvected with the geodesic null vectors tangent to the  null generators of ${\cal N}_p$ imply then that 
\[
\Omega = 0 
\quad \mbox{and} \quad  \Pi  \neq 0
\quad \mbox{on} \quad {\cal N}_p, \quad \quad  
\nabla_{j}\Omega \neq  0 \quad \mbox{on} \quad {\cal N}_{p} \setminus \{p\}.
\]
(Note that the assumption $ \Pi |_{p} = 0$ would  imply  that $\nabla_j\Omega =  0$
on ${\cal N}_{p}$).

The sign of $ \Pi $ depends on the signature of $g$. The equation 
$\nabla_{\mu}\,\nabla_{\nu}\,\Omega = - \Omega\,L_{\mu \nu} +  \Pi \,g_{\mu \nu}$
implies for a future directed time-like geodesics $\gamma$  starting at $p$ the relation 
$ \Pi \,g(\gamma', \gamma')|_p = \nabla_{\gamma'} \nabla_{\gamma'}\Omega|_p $. If we want this to be positive we must assume that $sign( \Pi ) = sign(g(\gamma', \gamma')) = sign(\eta_{00})$ at $i^-$. 
This discussion shows that with the assumptions above on $\Omega$ and $ \Pi $ at $p$ the field equations themselves will take care for the conformal factor $\Omega$ to evolve so that it will show near $p$  the desired behaviour on ${\cal N}_p$ and on the physical space-time region $I^+({\cal N}_{p})$.

\vspace{.3cm}

Under a rescaling $g_{\mu \nu} \rightarrow \hat{g}_{\mu \nu} = \theta^2\,g_{\mu \nu}$, 
$\Omega \rightarrow \hat{\Omega} = \theta\,\Omega$
with some function $\theta > 0$ it follows
\[
 \Pi |_{p}  \rightarrow \hat{ \Pi }|_{p} =  (\Pi\,\theta^{-1}) |_p.
\]
The transformation laws
\[
R_{\mu \nu}[g] \rightarrow R_{\mu \nu}[\hat{g}] = R_{\mu \nu}[g]  
- 2\,\theta^{-1}\,\nabla_{\mu}\,\nabla_{\nu}\,\theta 
+ 4\,\theta^{-2}\,\nabla_{\mu}\,\theta\,\nabla_{\nu}\,\theta 
\]
\[
- \{ \theta^{-1}\,\nabla_{\lambda}\,\nabla^{\lambda}\,\theta 
+ \theta^{-2}\,\nabla_{\lambda}\,\theta\,\nabla^{\lambda}\,\theta 
\}\,g_{\mu \nu},
\]
and 
\begin{equation}
\label{R-conf-transf-rule}
R[g] \rightarrow R[\hat{g}] = \theta^{-2}\,\{R[g] - 6\, \theta^{-1}\,\nabla_{\lambda}\,\nabla^{\lambda}\,\theta \},
\end{equation}
of the Ricci tensor and the Ricci scalar
imply the transformation behaviour 
\[
S_{\mu \nu}[g] \rightarrow S_{\mu \nu}[\hat{g}] = S_{\mu \nu}[g]  
- 2\,\theta^{-1}\,\nabla_{\mu}\,\nabla_{\nu}\,\theta 
+ 4\,\theta^{-2}\,\nabla_{\mu}\,\theta\,\nabla_{\nu}\,\theta 
\]
\[
+ \frac{1}{2}\, \{ \theta^{-1}\,\nabla_{\lambda}\,\nabla^{\lambda}\,\theta 
- 2\, \theta^{-2}\,\nabla_{\lambda}\,\theta\,\nabla^{\lambda}\,\theta 
\}\,g_{\mu \nu}.
\]
Let $l^{\mu} \neq 0$ denote the tangent vector of a future directed null geodesics $\gamma(\tau)$ 
on $ {\cal J}_{p}$ with $\gamma(0) = p$, so that $\nabla_l l = 0$. Then 
$\hat{l} = \theta^{-2}\,l$ satisfies $\hat{g}(\hat{l}, \hat{l}) = 0$, 
$\hat{\nabla}_{\hat{l}} \hat{l} = 0$. This gives
\[
\theta^4\,\hat{l}^{\mu}\,\hat{l}^{\nu}\,S_{\mu \nu}[\hat{g}] = 
l^{\mu}\,l^{\nu}S_{\mu \nu}[g]  
- 2\,\theta^{-1}\,(l^{\mu}\nabla_{\mu})^2\,\theta 
+ 4\,\theta^{-2}\,(l^{\mu}\,\nabla_{\mu}\,\theta)^2, 
\]
or equivalently
\begin{equation}
\label{theta-ODE}
\hat{l}^{\mu}\,\hat{l}^{\nu}\,S_{\mu \nu}[\hat{g}]\,\,\theta^3 = 
l^{\mu}\,l^{\nu}\,S_{\mu \nu}[g] \,\,\theta^{-1} 
+ 2\,(l^{\mu}\nabla_{\mu})^2\,(\theta^{-1}).
\end{equation}
For prescribed value of $\hat{l}^{\mu}\,\hat{l}^{\nu}\,S_{\mu \nu}[\hat{g}]$ this represents an ODE for $\theta$ along the null generator tangent to $l$. While the value of $\theta$ can be fixed at $p$ by specifying there the value of $| \Pi |$, there remains the freedom to specify the value of $\nabla_{\mu}\theta$ at $p$.
The equations above suggest  that a convenient  conformal gauge can be defined 
in a neighbourhood of $p$ in $J^+({\cal N}_{p})$
by requiring 
 \begin{equation}
 \label{A-conf-gauge}
 \Omega = \,0, \,\,\,\,\nabla_{\mu}\Omega = \,0,\,\,\,\,
  \Pi  = \,2\,\eta_{00}\, \,\,\,\mbox{at}\,\,\,\,p,
\end{equation}
and
 \begin{equation}
 \label{B-conf-gauge}
 \quad l^{\mu}\,l^{\nu}\,S_{\mu \nu}[g] = 0\,\,\,
 \mbox{on}\,\,\,{\cal N}_{p}\,\,\,\mbox{near $p$}, \quad R[g] = 0 \,\,\,
\mbox{on $J^+({\cal N}_{p})$ near $p$}.
\end{equation}
This conformal gauge will be assumed in the following without any problem. When this  type of conformal gauge  is used in a wider context, however, it is important to know  that for a {\it given} smooth background $g$ equation  (\ref{theta-ODE}) with 
$\hat{l}^{\mu}\,\hat{l}^{\nu}\,S_{\mu \nu}[\hat{g}] = 0$ 
yields a rescaling factor $\theta$ on ${\cal N}_p$ which has the appropriate smoothness behaviour on ${\cal N}_p$ near the vertex $p$ so that the wave equation obtained on the right hand side of (\ref{R-conf-transf-rule})
by setting $R[\hat{g}]  = 0$ can be solved with these data on ${\cal N}_p$ for a smooth function $\theta$ near $p$.
This question will be discussed in the article \cite{chrusciel:paetz:2013}.

\vspace{.4cm}

\noindent
{\it The  choice of the coordinates near $i^-$. }

\vspace{.4cm}

We shall consider $p$-centered $g$-normal coordinates $x^{\mu}$ near $p$. 
These are determined by the requirements that 
$x^{\mu}(p) = 0$, that $g_{\mu \nu}(0) = \eta_{\mu \nu}$ and that  for  given $x^{\mu} \neq 0$ and a real parameter $\tau$ with $|\tau|$ small enough the curve $\gamma: \tau \rightarrow \tau\,x^{\mu}$
is a geodesic through the point $p$. If  $g_{\mu \nu}$ and
$\Gamma_{\mu}\,^{\rho}\,_{\nu}$ denote the metric coefficients and the Christoffel symbols in the coordinates $x^{\mu}$ the latter condition is equivalent to
\[
0 = 2\,g_{\mu \rho}\,(\nabla_{\gamma'} \gamma')^{\rho} 
= 2\,g_{\mu \rho}\,x^{\nu}\,\Gamma_{\nu}\,^{\rho}\,_{\lambda}(\tau\,x)\,x^{\lambda}
= 2\,x^{\nu}\,g_{\nu \mu, \lambda}(\tau\,x)\,x^{\lambda}
- x^{\nu}\,x^{\lambda}\,g_{\nu \lambda, \mu}(\tau, x),
\]
which gives in particular that 
\begin{equation}
\label{normal-Christoffel-s}
x^{\nu}\,\Gamma_{\nu}\,^{\rho}\,_{\lambda}(\tau\,x)\,x^{\lambda} = 0,
\end{equation}
for small enough $|\tau|$. The first equation above implies further
$0 = x^{\nu}\,x^{\mu}\,g_{\nu \mu, \lambda}(\tau\,x)\,x^{\lambda}
= \frac{d}{d\tau}\left(x^{\nu}\,x^{\mu}\,g_{\nu \mu}(\tau\,x)\right)$
and thus 
$x^{\nu}\,x^{\mu}\,g_{\nu \mu}(\tau\,x) = x^{\nu}\,x^{\mu}\,g_{\nu \mu}(0)$,
whence
\[
2\,x^{\nu}\,g_{\nu \mu}(\tau\,x) 
+ \tau\,x^{\nu}\,x^{\lambda}\,g_{\nu \lambda, \mu}(\tau\,x) 
= 2\,x^{\nu}\,g_{\nu \mu}(0).
\]
With the first equations it follows  then
\[
0 = \tau\,2\,x^{\nu}\,g_{\nu \mu, \lambda}(\tau\,x)\,x^{\lambda}
- \tau\,x^{\nu}\,x^{\lambda}\,g_{\nu \lambda, \mu}(\tau, x)
= 2\,\frac{d}{d\tau} \left\{ \tau \left( x^{\nu}\,g_{\nu \mu}(\tau\,x)
- x^{\nu}\,g_{\nu \mu}(0) \right) \right \},
\]
and thus 
\begin{equation}
\label{norm-coord-char}
x^{\nu}\,g_{\nu \mu}(\tau\,x) = x^{\nu}\,g_{\nu \mu}(0).
\end{equation}

\vspace{.1cm}

\noindent
This equation implies in turn 
$x^{\nu}\,x^{\mu}\,g_{\nu \mu}(\tau\,x) = x^{\nu}\,x^{\mu}\,g_{\nu \mu}(0)$ which gives by differentiation 
$\tau\,x^{\nu}\,x^{\lambda}\,g_{\nu \lambda, \mu}(\tau\,x) 
= - 2\,x^{\nu}\,g_{\nu \mu}(\tau\,x)  + 2\,x^{\nu}\,g_{\nu \mu}(0) = 0$.
Because differentiation of  (\ref{norm-coord-char}) with respect to $\tau$ gives
$0 =  x^{\nu}\,g_{\nu \mu, \lambda}(\tau\,x)\,x^{\lambda}$ we see that 
 (\ref{norm-coord-char}) implies that the curves $\gamma$ considered above are in fact geodesics. The relation (\ref{norm-coord-char}) thus completely characterizes normal coordinates in terms of algebraic conditions on  the metric coefficients.
It follows from the equations above that
$g_{\mu \nu,\,\rho}(p) = 0$, $\Gamma_{\mu}\,^{\rho}\,_{\nu}(p) = 0$.\\

\noindent
In this gauge  ${\cal N}_p$ is now given by the set $\{x^{\mu} \in \mathbb{R}^4|\,\,
\eta_{\mu \nu}\,x^{\mu}\,x^{\nu},\,\,x^0 \ge 0\}$.

\vspace{.4cm}

\noindent
{\it The choice of the frame near $i^-$.}

\vspace{.4cm}

Assume now that $p$-centered  $g$-normal coordinates $x^{\mu}$ are given on a convex normal neighbourhood $U'$ of $p$ and take their values in a neighbourhood $U$ of the origin of $\mathbb{R}^4$. A frame $\{e_k\}_{k = 0, 1,2,3}$ is called a {\it normal frame centered at 
$p$}  if it satisfies on $U'$
\[
g(e_j,\,e_k) = \eta_{jk}, \quad \mbox{and} \quad \nabla_{\gamma'}e_k = 0,
 \]
for any geodesic $\gamma$ passing through $p$. The frame coefficients  satisfying $e_k = e^{\mu}\,_k\,\partial_{\mu}$ 
are assumed to satisfy
\[
e^{\mu}\,_k(0) = \delta^{\mu}_k.
\]
The 1-forms dual to $e_k$ will be denoted by  $\sigma^j$.
Then $\sigma^j = \sigma^j\,_{\nu}\,dx^{\nu}$ 
with $\sigma^j\,_{\mu}\, e^{\mu}\,_k = \delta^j_k$.
That the frame field depends in fact smoothly on the coordinates $x^{\mu}$ follows by  arguments known from the discussion of the exponential function.

The equation $x^{\nu}\,g_{\nu \mu}(\tau\,x) \,e^{\mu}\,_k(\tau\,x) = 
x^{\nu}\,\eta_{\nu \mu}\,\delta^{\mu}_k$ expresses that
the scalar product  $g(\gamma', e_k)$ is constant along the geodesic $\gamma$. The representation $g_{\mu \nu} = \eta_{ij}\,\sigma^i\,_{\mu}\,\sigma^j\,_{\nu}$ allows us to rewrite it in the form
\begin{equation}
\label{x-sigma}
x^{\mu}\,\sigma^j\,_{\mu}(\tau\,x) = x^{\mu}\,\delta^j_{\mu} 
\quad \mbox{resp.}\quad
x^{\mu}\,\delta^j_{\mu}\,e^{\nu}\,_j(\tau\,x) = x^{\nu}.  
\end{equation}
With this relation  equation  (\ref{norm-coord-char}) implies
\begin{equation}
\label{sigma-x}
x^{\mu}\,\eta_{\mu \rho}\,\delta^{\rho}_j\,\sigma^j\,_{\nu}(\tau\,x) = x^{\mu}\,\eta_{\mu \nu}
\quad \mbox{resp.}\quad x^{\mu}\,\eta_{\mu \nu}\,e^{\nu}\,_k(\tau\,x) = 
 x^{\mu}\,\eta_{\mu \nu}\,\delta^{\nu}\,_k.
\end{equation}

If the fields $\sigma^j\,_{\mu}$ and the coordinates $x^{\mu}$  satisfy the last two relations it follows  without further assumptions that the metric $g_{\mu \nu} = \eta_{ij}\,\sigma^i\,_{\mu}\,\sigma^j\,_{\nu}$ satisfies (\ref{norm-coord-char}). In terms of the frame field the information that the $x^{\mu}$ are normal coordinates is thus encoded in  (\ref{x-sigma}), (\ref{sigma-x}).

\vspace{.1cm}

Writing $\nabla_i \equiv \nabla_{e_i}$, the connection coefficients $\Gamma_i\,^j\,_k$ with respect to the frame $e_j$ are defined by the relations $\nabla_i\,e_k = \Gamma_i\,^j\,_k\,e_j$. They satisfy $\Gamma_{ijk} = - \Gamma_{ikj}$, where $\Gamma_{ijk} =
 \Gamma_i\,^l\,_k\,\eta_{lj}$.

\vspace{.1cm}

The tensor field $X(x) = x^{\mu}\,\partial_{\mu}$  tangential to the geodesics through $p$ is characterized uniquely by the conditions
\begin{equation}
\label{X-definition}
X(p) = 0, \quad \nabla_{\mu}\,X^{\nu}(p) = g_{\mu}\,^{\nu}(p), \quad \nabla_X X = X.
\end{equation}
By (\ref{x-sigma}) it can be written   $X = X^k e_k$ with $X^k(x) = \delta^k_{\nu}\,x^{\nu}$.
The relation $\nabla_X e_j = 0$ is equivalent to
\begin{equation}
\label{nablaX-e=0}
X^k(x)\,\Gamma_k\,^i\,_j(x) = \delta^k_{\nu}\,x^{\nu}\,\Gamma_k\,^i\,_j(x) = 0, 
\quad x^{\mu} \in U,
\end{equation}
or 
\begin{equation}
\label{spin-nablaX-e=0}
X^{AA'}(x)\,\Gamma_{AA'}\,^B\,_C(x) = 0, 
\quad x^{\mu} \in U.
\end{equation}
This is the characterizing  property of the normal frame.

\vspace{.1cm}

In the following we shall refer to coordinates $x^{\mu}$ and a frame $e_k$ (resp. 
$e_{AA'}$) which satisfy the conditions  above as  to a {\it normal gauge}. We shall always assume this to be supplemented by a normalized spin-frame $\{\iota_A\}_{A = 0, 1} $ which satisfies $e_{AA'} = \iota_A\,\bar{\iota}_{A'}$ and $\nabla_X \iota_A = 0$. All spinor fields will be assumed to be given in this frame.

\section{Normal expansions}
\label{null-data}
\vspace{.5cm}

Let $x^{\mu}$ and $e_{AA'}$ be given in a normal gauge and let $X$ be the vector field defined by (\ref{X-definition}) so that $X = X^i\,e_i = X^{AA'}\,e_{AA'}$ with 
$X^{AA'}(x) = x^{\mu}\,\alpha_{\mu}\,^{AA'}$,
where we set $\alpha_{\mu}^{AA'} = \delta^i\,_{\mu}\,\alpha_{i}\,^{AA'}$.

Let  $T$ denote a smooth spinor  field and $T_{A_1 \ldots A_j B'_1 \ldots B'_k}$
its components  in  the normal frame. 
If $x^{\mu}_* \neq 0$, then we get with 
(\ref{x-sigma}) and (\ref{spin-nablaX-e=0}) along the geodesic 
$\gamma: \tau \rightarrow \tau\,x^{\mu}_*$
\[
\frac{d}{d\tau}T_{A_1 \ldots A_j B'_1 \ldots B'_k}(\tau\,x_*) =
T_{A_1 \ldots A_j B'_1 \ldots B'_k, \,\mu}(\tau\,x_*)\,x^{\mu}_*
\]
\[
= x^{CC'}_* \left\{
T_{A_1 \ldots A_j B'_1 \ldots B'_k, \,\mu}(\tau\,x_*)\,e^{\mu}\,_{CC'}(\tau\,x_*)
\right.
\]
\[
\left.
- \Gamma_{CC'}\,^{D_1}\,_{A_1}\,T_{D_1 \ldots A_j B'_1 \ldots B'_k}(\tau\,x_*) 
\ldots 
-  \bar{\Gamma}_{CC'}\,^{E'_k}\,_{B'_k}\,T_{A_1 \ldots A_j B'_1 \ldots B'_k}(\tau\,x_*) \right\}
\]
\[
 = x^{CC'}_*
\nabla_{CC'}T_{A_1 \ldots A_j B'_1 \ldots B'_k}(\tau\,x_*)
\]
with $x^{CC'}_* =  x^{\mu}_*\,\delta^i\,_{\mu}\,\alpha_i\,^{CC'}$. Applying the argument repeatedly gives
\[
\frac{d^n}{d\tau^n}T_{A_1 \ldots A_j B'_1 \ldots B'_k}(\tau\,x_*) 
=
 x^{C_1C'_1}_* \ldots  x^{C_nC'_n}_*
\nabla_{C_1C'_1} \ldots \nabla_{C_nC'_n}T_{A_1 \ldots A_j B'_1 \ldots B'_k}(\tau\,x_*).
\]
Setting  $x^{\mu} = \tau\,x^{\mu}_*$ in the Taylor expansion
\[
T_{A_1 \ldots A_j B'_1 \ldots B'_k}(\tau\,x_*)  =
\sum_{n = 0}^N \frac{1}{n!} \,\tau^n \,\frac{d^n}{d\tau^n}T_{A_1 \ldots A_j B'_1 \ldots B'_k}(0)\,
+ O(|\tau|^{N + 1}), 
\]
the Taylor expansion of $T_{A_1 \ldots A_j B'_1 \ldots B'_k}$ at $p$ is obtained in the form

\begin{equation}
\label{normal-expansion}
T_{A_1 \ldots A_j B'_1 \ldots B'_k}(x)  
= \sum_{{|{\bf n}|} = 0}^N \frac{1}{|{\bf n}| !}\, 
 X^{\bf n}\,\nabla_{\bf n}T_{A_1 \ldots A_j B'_1 \ldots B'_k}(0)
+ O(|x|^{N + 1}) 
\end{equation}
\[
= \sum_{{|{\bf n}|} = 0}^N \frac{1}{|{\bf n}| !}\, 
 X^{\bf n}\,\nabla_{(\bf n)}T_{A_1 \ldots A_j B'_1 \ldots B'_k}(0)
+ O(|x|^{N + 1}). 
\]
This will be  referred to as the {\it normal expansion} of $T$ at $p$.
It   will be known once the {\it symmetrized} covariant derivatives
$\nabla_{(\bf n)}\,T_{A_1 \ldots A_j B'_1 \ldots B'_k}(p)$, $\,|{\bf n}|  > 0$,
are given.

\subsection{The null data.}

\vspace{.4cm}

The set $C_p \sim S^2$ of future directed 
null vectors at $p$ satisfying $g(l, l) = 0$ and  $g(l, e_0) = \eta_{00}/\sqrt{2}$
defines a parametrization of the null generators of ${\cal N}_p$, which are given in the normal gauge by the curves $\tau \rightarrow \tau\,l^{\mu}$,  $l^{\mu} \in C_p$, $0 \le \tau < a$ for some suitable $a > 0$. Denote by  ${\cal W}_p$  the subset of ${\cal N}_p$ which is  generated by the null generators parametrized by a proper  open subset $W$ of $C_p$. 

Let $\kappa^A(x) $ be a smooth spinor field on ${\cal W}_p \setminus \{p\}$ which is parallely propagated along the null generators and such that 
$\kappa^A \,\bar{\kappa}^{A'}$ is tangent to the null generators of ${\cal W}_p$.
Because the components $\kappa^A$ are given in the normal frame they are constant  along the null generators. 
Thus, $\kappa^A$ assumes a limit as $\tau \rightarrow 0$ along the curve 
$\tau \rightarrow \tau\,l^{\mu}$ and it can be assumed that
 $\kappa^A \,\bar{\kappa}^{A'} = l^{AA'}$ along that curve.
The field $\kappa^A$ is then determined uniquely up to phase transformations
$\kappa^A \rightarrow e^{i\,\phi}\,\kappa^A$ with smooth phase factors which are constant along the null generators.

For a given tensor field $T$ with spin frame components $T_{A_1 \,\ldots \,
A_j B'_1 \,\ldots \, B'_k}$ we define  its {\it null datum} on ${\cal W}_p$ as the spin weighted function 
\[
T_0(x) = \kappa^{A_1}(x) \ldots  
 \kappa^{A_j}(x) \bar{\kappa}^{B'_1}(x) \ldots \bar{\kappa}^{B'_k}(x) \,T_{A_1 \,\ldots \,
A_j B'_1 \,\ldots \, B'_k}(x), \quad x^{\mu} \in {\cal W}_p \setminus \{p\}.
\] 
With the normal expansion for $T$ given above this gives at $p$  the asymptotic representation 
\[
T_0(\tau\,x)  
=
\sum_{n = 0}^N \frac{\tau^n}{n!}\, 
 \kappa^{C_1} \,\ldots  \,\bar{\kappa}^{C'_n}
 \,\kappa^{A_1} \ldots \bar{\kappa}^{B'_k}\,
\nabla_{C_1C'_1}\, \ldots\, \nabla_{C_nC'_n}\,T_{A_1 \ldots A_j B'_1 \ldots B'_k}(0)
+ O(|\tau|^{N + 1}). 
\]
for $\tau > 0$. The sum is determined uniquely by the coefficients 
\[
\tilde{T}_n(\kappa) =    \kappa^{C_1}\,\dots \,\bar{\kappa}^{C'_n} 
 \,\kappa^{A_1} \,\ldots \,\bar{\kappa}^{B'_k}\,
\nabla_{C_1C'_1} \,\ldots \,\nabla_{C_nC'_n}\,T_{A_1 \ldots A_j B'_1 \ldots B'_k}(0).
\]
Because the directions $\kappa^A \,\bar{\kappa}^{A'} = l^{AA'}$ are allowed to vary in the open subset $W$ of $C_p$, knowing these coefficients is equivalent to knowing 
the symmetrized derivatives
\begin{equation}
\label{A-data-jet}
T_{(A_1 \ldots A_j)}\,^{(B'_1 \ldots B'_k)}(0), \quad 
\nabla_{(C_1}\,^{(C'_1} \ldots \nabla_{C_n}\,^{C'_n}\,T_{A_1 \ldots A_j)}\,^{B'_1 \ldots B'_k)}(0), \quad n = 1, 2, \ldots\,\,.
\end{equation}
In fact, let $S_{A_1 \ldots A_p\,A'_1 \ldots A'_q} = S_{(A_1 \ldots A_p)\,(A'_1 \ldots A'_q)}$ be a symmetric spinor.  It will be known once its  `essential components', denoted by 
 $S_{ij} = S_{(A_1 \ldots A_p)_i\,(A'_1 \ldots A'_q)_j}$, are known, which are obtained by 
setting  for  given integers $i, j$, with $0 \le i \le  p$, $0 \le j \le  q$, precisely $i$ unprimed resp. $j$ primed indices to equal to one. Choose 
$(\kappa^0, \kappa^1) = \beta\,(1, z)$ with $z \in \mathbb{C}$ and the factor 
$\beta = (1 + |z|^2)^{- 1/2}$ which ensures the normalization condition on $l^{\mu}$.
If the function 
$S(\kappa) = \kappa^{A_1} \ldots \bar{\kappa}^{A'_q}\,S_{A_1 \ldots A_p\,A'_1 \ldots A'_q}$ 
is known then also the  function 
\[
\beta^{-p- q}\,S(\kappa) = \sum_{i = 0}^p\,\sum_{j = 0}^q
{p \choose i}{q \choose j}\,S_{ij}\,z^i\,\bar{z}^j,
\]
and the essential components are given by  $S_{ij} 
= \frac{(p - i)!}{p!}\,\frac{(q - j)!}{q!}\,\partial_z^i\,\partial_{\bar{z}}^j(\beta^{-p -q}S(\kappa))|_{z = 0}$.

\vspace{.3cm}

While the {\it null datum on ${\cal W}_p$}  is a spin weighted function which depends on the choice of $\kappa^A$, the spinors 
(\ref{A-data-jet}) at $p$ are given with respect to the spin-frame $\iota_A$
and are  independent of any phase factors.
They will be referred to as to the {\it null data  of $T$ at $p$}.

\vspace{.3cm}

Of particular importance will be for us the null datum 
\begin{equation}
\label{rad-field}
\psi_0 = \kappa^A\,\kappa^B\,\kappa^C\,\kappa^D\,\psi_{ABCD},
\end{equation} 
associated with the rescaled conformal Weyl spinor $\psi_{ABCD}$. It   is referred to as the 
{\it radiation field}. 

To illustrate some of its properties it will be convenient to proceed as follows.
Let $SU(2, \mathbb{C})$ denote the subgroup of transformations $(s^A\,_B)_{A, B = 0, 1} \in sl(2, \mathbb{C})$ which satisfy
$\epsilon_{AC}\,s^A\,_B\,s^C\,_D = \epsilon_{BD}$
and $s^A\,_B\,\bar{s}^{A'}\,_{B'}\,\alpha_0^{BB'} = \alpha_0^{AA'}$. Then  the null vectors $l^{\mu} = l^{\mu}(s)$ at $p$ with spinor components $l^{AA'} = s^A\,_0\,\bar{s}^{A'}\,_{0'}$ sweep out the null directions at $p$ and the $m^{AA'} = m^{AA'}(s) = s^A\,_0\,\bar{s}^{A'}\,_{1'}$ are complex null vectors orthogonal to $l^{AA'}$. By requiring them to be constant along the null generators tangent to $l^{AA'}$ they will be parallely transported and tangent to ${\cal N}_p$ along the generators.

The information on the radiation field is equivalent to the information contained in the pull back of the tensor $W_{ijkl}\,l^i\,l^k$ to ${\cal N}_p$. In fact, the latter can be specified  by the contractions 
of the symmetric tensor $W_{ijkl}\,l^i\,l^k$ with the field $m$ and $\bar{m}$. Because
$W_{ijkl}\,l^i\,m^j\,l^k\,m^l$ and $W_{ijkl}\,l^i\,\bar{m}^j\,l^k\,\bar{m}^l$ are complex conjugates of each other and the trace-freeness of  $W_{ijkl}$ implies that $W_{ijkl}\,l^i\,m^j\,l^k\,\bar{m}^l = 0$, the information  
is stored in $W_{ijkl}\,l^i\,m^j\,l^k\,m^l = s^A\,_0\,s^B\,_0\,s^C\,_0\,s^D\,_0\,\psi_{ABCD}
= \psi_0$. Note that this description includes the complete freedom to perform phase transformations. If this is to be removed, one has to restrict the choice of $s$ to a local section of the Hopf map
$SU(2) \ni s \rightarrow l^{AA'}(s) \in S^2$, where $S^2$ is identified with the set of future directed null directions at $p$.

The {\it null data of $\psi$ at $p$} can be  extracted from the {\it null datum $\psi$ on ${\cal N}_p$} as follows. 
By taking derivatives with respect to $\tau$
at $\tau = 0$ one gets from the null datum  the quantities 
\[
\tilde{\psi}_n(s) = s^{C_1}\,_0\,\bar{s}^{C'_1}\,_{0'} \ldots s^{C_n}\,_0\,\bar{s}^{C'_n}\,_{0'}
s^A\,_0\,s^B\,_0\,s^C\,_0\,s^D\,_0\,\nabla_{C_1C'_1} \ldots \nabla_{C_nC'_n}\psi_{ABCD}(0).
\]
As discussed in detail in \cite{friedrich:pure-rad:1986}, these functions on $SU(2, \mathbb{C})$ translate naturally  into expansions in terms of the coefficients $T_m\,^i\,_j(s)$
of certain finite unitary representations of the group  
$SU(2, \mathbb{C})$. With this understanding the essential components of the null data 
$\nabla_{(C_1}\,^{(C'_1} \ldots \nabla_{C_n}\,^{C'_n)}\psi_{ABCD)}(0)$ can be obtained by performing integrals of $\tilde{\psi}_n(s)\,\bar{T}_m\,^i\,_j(s)$ with respect to the Haar measure on $SU(2, \mathbb{C})$.
Any ambiguities related  to choices of phase factors as indicated above are cancelled out by the integration.

\vspace{.2cm}

To prescribe the null datum in a way  which ensures the necessary smoothness properties we start with some symmetric spinor field $\psi^*_{ABCD} = \psi^*_{ABCD}(x^{\mu})$ which is 
defined and smooth in a suitable neighbourhood of the origin $p$ of  $\mathbb{R}^4$ (so that $x^{\mu}(p) = 0$). This field will be thought as being  given in a conformal and normal  gauge as described in section \ref{gauge-cond}. 
Assuming  $s^A\,_B$ as above, one can then consider
on the cone 
${\cal N}_p = \{\eta_{\mu \nu}\,x^{\mu}\,x^{\nu} = 0, \,\,x^0 \ge 0\}$ (or more precisely on the bundle $\tilde{\cal N}_p \sim \mathbb{R}^+_0 \times SU(2)$ over ${\cal N}_p$, see section
\ref{transport-equ-and-inner-constr})
the complex-valued function
\begin{equation}
\label{smooth-radiation-field}
\psi_0(\tau, s) = 
s^A\,_0\,s^B\,_0\,s^C\,_0\,s^D\,_0\,
\psi^*_{ABCD}(\tau\,\alpha^{\mu}_{EE'}\,s^E\,_0\,\bar{s}^{E'}\,_{0'}),
\end{equation}
as a `smooth' radiation field.

\vspace{.2cm}

The gauge conditions  give  control on the null data at $p$ for some of the unknowns in the conformal field equations.
 It follows immediately from the discussion above and the first of conditions 
(\ref{B-conf-gauge}) that the conformal gauge implies 
\begin{equation}
\label{B2-conf-gauge}
\Phi_{A B}\,^{A' B'}(0) = 0, \quad 
\nabla_{(C_1}\,^{(C'_1} \ldots \nabla_{C_n}\,^{C'_n}\,\Phi_{A B)}\,^{A' B')}(0) = 0, 
\quad n = 1, 2, \ldots\,\,.
\end{equation}

\section{Formal expansions at $i^-$.}
\label{formal expansion}
\vspace{.2cm}

In a conformal gauge satisfying (\ref{B-conf-gauge}) the conformal field equations  read

\begin{equation}
\label{Omega-equ}
\nabla_{AA'}\,\nabla_{BB'}\,\Omega = - \Omega\,\Phi_{AB A' B'} +  \Pi \,\epsilon_{AB}\,\epsilon_{A'B'},
\end{equation}
\begin{equation}
\label{s-equ}
\nabla_{AA'}\, \Pi  = - \nabla^{BB'}\Omega\,\Phi_{AB A'B'},
\end{equation}
\begin{equation}
\label{Phi-equ}
\nabla_A\,^{D'}\,\Phi_{BCB'D'} 
= \psi_{ABCD}\,\nabla^D\,_{B'}\Omega,
\end{equation}
\begin{equation}
\label{psi-equ}
 \nabla^D\,_{B'}\,\psi_{ABCD} = 0,
\end{equation}
and the curvature spinor (\ref{spin-curvature-decomposition}) takes the form
\begin{equation}
\label{gauged-spin-curvature-decomposition}
R_{ABCC'DD'} = \Omega\,\psi_{ABCD}\,\epsilon_{C'D'} + \Phi_{ABC'D'}\,\epsilon_{CD}.
\end{equation}

\vspace{.1cm}

\noindent
The following algebraic considerations will be simplified by rewriting equations
(\ref{Phi-equ}) and (\ref{psi-equ}).
The symmetry of $\psi_{ABCD}$ and the fact that  vanishing spinor contractions indicate index  symmetries imply that equation (\ref{psi-equ}) is equivalent to 
\begin{equation}
\label{2-Bianchi-equ}
\nabla_{E}\,^{E'}\,\psi_{ABCD} = \nabla_{(E}\,^{E'}\,\psi_{ABCD)}.
\end{equation}
If (\ref{psi-equ}) holds, equation (\ref{Phi-equ})
and its complex conjugate are equivalent to the equations 
\begin{equation}
\label{unprimed-dPhi-ind-com}
\nabla_A\,^{A'}\,\Phi_{BC}\,^{B'C'} -
 \nabla_B\,^{A'}\,\Phi_{AC}\,^{B'C'} = - \,\epsilon_{AB}\,\nabla_C\,^{H'}\Omega\,\,
 \bar{\psi}^{A'B'C'}\,_{H'},
\end{equation}
\begin{equation}
\label{primed-dPhi-ind-com}
\nabla_A\,^{A'}\,\Phi_{BC}\,^{B'C'} - 
 \nabla_A\,^{B'}\,\Phi_{BC}\,^{A'C'} = - \,\epsilon^{A'B'}\,\nabla^{H C'}\Omega\,\,
 \psi_{ABCH}.\quad
 \end{equation}
With the identity 
\[
\nabla_A\,^{A'}\,\Phi_{BC}\,^{B'C'} = \nabla_{(A}\,^{(A'}\,\Phi_{BC)}\,^{B'C')} 
\]
\[
+ \frac{2}{3}\,\nabla_{(A}\,^{H'}\,\Phi_{BC)H'}\,^{(B'}\,\epsilon^{C')A'}
-  \frac{2}{3}\,\epsilon_{A(B}\,\nabla^{H(A'}\,\Phi_{C)H}\,^{B'C')}
-  \frac{4}{9}\,\epsilon_{A(B}\,\nabla^{HH'}\,\Phi_{C)H H'}\,^{(B'}\,\epsilon^{C')A'},
\]
these two equations are seen to be equivalent to the equation
\begin{equation}
\label{2-Bianchi-id+Bianchi-equ}
\nabla_A\,^{A'}\,\Phi_{BC}\,^{B'C'} = \nabla_{(A}\,^{(A'}\,\Phi_{BC)}\,^{B'C')} 
\end{equation}
\[
+ \frac{2}{3}\,\psi_{ABCH}\,\nabla^{H(B'}\,\Omega\,\,\epsilon^{C')A'}
+ \frac{2}{3}\,\epsilon_{A(B}\,\nabla_{C)H'}\Omega\,\,\bar{\psi}^{A'B'C'H'}.
\]

\vspace{.5cm}

\noindent
We note that
\begin{equation}
\label{psi-0-1-data}
\psi_{ABCD}(0), \quad \quad
\nabla_{E}\,^{E'}\,\psi_{ABCD}(0) = \nabla_{(E}\,^{E'}\,\psi_{ABCD)}(0),
\end{equation}
represent null data of $\psi_{ABCD}$ and that the conformal gauge 
(\ref{A-conf-gauge}), (\ref{B-conf-gauge}) 
implies by  (\ref{B2-conf-gauge}) and (\ref{2-Bianchi-id+Bianchi-equ}) that
\begin{equation}
\label{Phi-0-1-data}
\Phi_{BC}\,^{B'C'}(0) = 0,\quad 
\nabla_A\,^{A'}\,\Phi_{BC}\,^{B'C'}(0)  = \nabla_{(A}\,^{(A'}\,\Phi_{BC)}\,^{B'C')}(0) = 0. 
\end{equation}
With this  it follows from equations 
(\ref{s-equ}), (\ref{Omega-equ}) and the gauge conditions that
\begin{equation}
\label{Omega-0 - - 5-data}
\nabla_{AA'}\nabla_{BB'}\Omega(0) =  \Pi (0)\,\epsilon_{AB}\,\epsilon_{A'B'}, 
\quad \nabla_{\bf k}\,\Omega(0) = 0 \quad \mbox{for} \quad |{\bf k}| = 0, 1, 3, 4, 5, 
\end{equation}
\begin{equation}
\label{s-0 -- 3 - data}
\nabla_{\bf k}  \Pi (0) = 0  \quad \mbox{for} \quad |{\bf k}| = 1, 2, 3.
\end{equation}
The relations above imply furthermore that 
\begin{equation}
\label{R=0x2}
R_{ABCC'DD'}(0) = 0, \quad \quad \nabla_{EE'}R_{ABCC'DD'}(0) = 0.
\end{equation}

\vspace{.2cm}
 
The following result, which relates the formal expansion of the curvature fields at a given point $p$ to the null data of $\psi_{ABCD}$ at $p$, applies and  extends arguments  of the theory of {\it exact sets of fields} discussed in \cite{penrose:1980}, \cite{penrose:rindler:I}.

\begin{lemma}
\label{unique-expansion}
In a neighbourhood of the point $p$ 
let the fields $\,\Omega$, $ \Pi $, $\Phi_{ABA'B'}$, $\psi_{ABCD}$, $e^{\mu}\,_{AA'}$, 
$\Gamma_{AA'}\,^B\,_C$  be smooth and be given  in a $p$-centered normal gauge for the coordinates and the frame and  in  a conformal gauge satisfying (\ref{A-conf-gauge}), (\ref{B-conf-gauge}). Then, if they satisfy the structural equations and the conformal field equations
the covariant derivatives  of  the fields $\,\Omega$, $ \Pi $, $\Phi_{ABA'B'}$, 
$\psi_{ABCD}$  at all orders  are determined uniquely  at $p$ 
by  the null data $\nabla_{(E_1}\,^{(E'_1} \ldots \nabla_{E_n}\,^{E'_n)}\,\psi_{ABCD)}(p)$, $n \in \mathbb{N}_0$, at $p$.\\ 

The resulting map which relates to the null data of $\psi$ at $p$ the covariant derivatives  of the fields $\,\Omega$, $ \Pi $, $\Phi_{ABA'B'}$, $\psi_{ABCD}$ at $p$ extends in a unique way so that it associates with  any freely  specified sequence of totally symmetric spinors 
\[
\xi_{ABCD}, \quad \xi_{E_1 \ldots  E_nABCD}^{E'_1 \ldots E_n'}, \quad n = 1, 2, 3, \ldots
\]
at $p$ formally `covariant derivatives' the of fields $\,\Omega$, $ \Pi $, 
$\Phi_{ABA'B'}$, 
$\psi_{ABCD}$  of any order  at $p$ such that
\begin{equation}
\label{data-jet-cond}
\psi_{ABCD}(p) = \xi_{ABCD}, \quad \nabla_{(E_1}\,^{(E'_1} \ldots \nabla_{E_n}\,^{E'_n)}\,\psi_{ABCD)}(p)
= \xi_{E_1 \ldots  E_nABCD}^{E'_1 \ldots E_n'}.
\end{equation}
\end{lemma}

\noindent
{\bf Remark}: The coefficients $e^{\mu}\,_{AA'}$ and $\Gamma_{AA'}\,^B\,_C$ have been listened  in the first statement because the field equations involve covariant derivatives of  tensor fields and thus require the frame and connection coefficients for their formulation. 
The following argument will, however, never make use of explicit expressions of covariant derivatives in terms of these coefficients and partial derivatives of the fields. It only  uses formal expressions of covariant derivatives and the standard rules for covariant derivatives such as commutation relations  and the Leibniz rule. Therefore the coefficients are not mentioned in the second part of the Lemma. How they are determined  will be discussed in the following section.

\vspace{.2cm}

\noindent
{\bf Proof}: 
At lowest order the first assertion of the Lemma follows from 
(\ref{psi-0-1-data}), 
(\ref{Phi-0-1-data}), 
(\ref{Omega-0 - - 5-data}) 
and
(\ref{s-0 -- 3 - data}). 
That  it is true at higher orders will be shown by an  induction argument.
In this  we shall repeatedly make use of (\ref{spin-commutator}) and (\ref{spin-curvature-decomposition}) with $\Lambda = 0$. With the identity
\[
\nabla_{CC'}\nabla_{DD'} - \nabla_{DD'}\nabla_{CC'} =
\epsilon_{CD}\,\nabla_{H(C'}\,\nabla^H\,_{D')} + \epsilon_{C'D'}\,\nabla_{(C |H'|}\,\nabla_{D)}\,^{H'}, 
\]
it is seen that  (\ref{spin-commutator}) and its complex conjugate are with our assumptions equivalent to the relations
\[
\epsilon_{C'D'}\,\nabla_{(C}\,^{C'} \nabla_{D)}\,^{D'}\kappa_A = \Omega\,\psi_{ABCD}\,\kappa^B,  \quad \,\,\,\,
\epsilon_{C'D'}\,\nabla_{(C}\,^{C'} \nabla_{D)}\,^{D'}\bar{\kappa}_{A'} =  \Phi_{CDA'B'}\,\kappa^{B'},  \quad\quad\,
\]
\[
\epsilon^{CD}\,\nabla_{C}\,^{(C'} \nabla_{D}\,^{D')}\kappa_A = \Phi_{AB}\,^{C'D'}\,\kappa^B, \quad\quad\,\,\,
\epsilon^{CD}\,\nabla_{C}\,^{(C'} \nabla_{D}\,^{D')}\bar{\kappa}_{A'} = \Omega\,\bar{\psi}_{A'B'}\,^{C'D'}\,\bar{\kappa}^{B'}. \quad
\]

\vspace{.2cm}

While the induction argument is fairly obvious 
for  the fields $\Omega$, $ \Pi $, it is more involved in the case of $\Phi_{ABA'B'}$ and $\psi_{ABCD}$.
The following observations are important. 
Consider  the quantities $\nabla_{E_1}\,^{E'_1}\, \ldots \,\nabla_{E_n}\,^{E'_n}\,\psi_{ABCD}$
with $n \ge 2$. If   the covariant derivatives would commute it would  follow that
\begin{equation}
\label{d-p-psi-decomp}
\nabla_{E_1}\,^{E'_1}\, \ldots \,\nabla_{E_n}\,^{E'_n}\,\psi_{ABCD}
= \nabla_{(E_1}\,^{(E'_1}\, \ldots \,\nabla_{E_n}\,^{E'_n)}\,\psi_{ABCD)}.
\end{equation}
In fact,  any order of the upper indices can be achieved by commuting the covariant derivatives. If can be shown that the lower indices can be brought into any order without changing the position of  the upper indices, the assertion will follow. Consider, for instance, 
the index positions given on the left hand side of the equation above.
To interchange the indices $E_k$ and $A$ (say) 
we commute $\nabla_{E_k}\,^{E'_k}$ to the right until we can use (\ref{2-Bianchi-equ}) to swap $E_k$ and $A$, then  
we commute again to bring  $\nabla_{A}\,^{E'_k}$ back to the $k$-th position.  To show that  indices $E_k$, $E_j$ can be interchanged
we operate with $\nabla_{E_k}\,^{E'_k}$ as before to get $\nabla_{A}\,^{E'_k}$, then commute $\nabla_{E_j}\,^{E'_j}$ to the right and use  (\ref{2-Bianchi-equ}) again to get $\nabla_{E_k}\,^{E'_j}$, then  commute $\nabla_{A}\,^{E'_k}$   to the right to get 
$\nabla_{E_j}\,^{E'_k}$ by using again  (\ref{2-Bianchi-equ}). Finally, commute  $\nabla_{E_j}\,^{E'_k}$ and 
$\nabla_{E_k}\,^{E'_j}$   into the $k$-th and $j$-th position respectively so that the order of the upper indices remains unchanged.

If the covariant derivatives do not commute one can still operate as above but use 
(\ref{spin-commutator}) and (\ref{spin-curvature-decomposition}) with $\Lambda = 0$ each time we commute derivatives. 
By this procedure  the curvature spinor 
$R^A\,_{BCC'DD'}$ and its derivatives enter the expressions and  (\ref{d-p-psi-decomp}) is replaced by an equation of the form
\begin{equation}
\label{psi=sym+rest}
\nabla_{E_1}\,^{E'_1}\, \ldots \,\nabla_{E_n}\,^{E'_n}\,\psi_{ABCD}
= \nabla_{(E_1}\,^{(E'_1}\, \ldots \,\nabla_{E_n}\,^{E'_n)}\,\psi_{ABCD)} \,+\, \ldots,
\end{equation}
where the dots indicate terms which depend on the curvature tensor and its derivatives and thus via the field equations on  the fields 
$\Omega$, $s$, $\Phi_{ABA'B'}$, $\psi_{ABCD}$ and their covariant derivatives of
order $\le n - 2$. Restriction to $p$ then implies with the induction hypothesis that the
$\nabla_{\bf n}\,\psi_{ABCD}(0)$with $|{\bf n}| \ge 2$ can be expressed in terms of $s(0)$ and the null data  of $\psi_{ABCD}$ of order $ \le n$.

Using (\ref{unprimed-dPhi-ind-com}) and (\ref{primed-dPhi-ind-com})  to interchange unprimed as well as primed indices we conclude by similar arguments that for $n \ge 2$
\begin{equation}
\label{Phii=sym+rest}
\nabla_{E_1}\,^{E'_1}\, \ldots \, \nabla_{E_n}\,^{E'_n}\,\Phi_{BC}\,^{B'C'} 
= \nabla_{(E_1}\,^{(E'_1}\, \ldots \, \nabla_{E_n}\,^{E'_n}\,\Phi_{BC)}\,^{B'C')}  +\, \ldots,
\end{equation}
where the dots indicate the terms of order $\le n - 2$, which are generated by commutating covariant derivatives and 
the terms which arise from the right hand sides of equations 
(\ref{unprimed-dPhi-ind-com}) and (\ref{primed-dPhi-ind-com}). These terms and the commutators contain 
expressions $\nabla_{\bf k}\psi_{ABCD}$, $\nabla_{\bf j} \bar{\psi}_{A'B'C'D'}$ with 
$|{\bf k}|, |{\bf j}| \le n - 1$
and derivatives $\nabla_{\bf l}\,\Omega$ with $|{\bf l}| \le n$. Equation 
(\ref{Omega-equ}) allows us to express the latter in terms of  $\nabla_{\bf m}\,\Omega$, 
$\nabla_{\bf p}\,s$ and $\nabla_{\bf q}\,\Phi_{ABCD}$ with 
$|{\bf m}|, |{\bf p}|, |{\bf q}| \le n - 2$.
Restricting to $x^{\mu} = 0$ and observing that the right hand side of (\ref{unprimed-dPhi-ind-com}), (\ref{primed-dPhi-ind-com}) vanish at $p$, we conclude with 
our induction hypothesis   that 
$\nabla_{\bf n}\,\Phi_{BCB'C'}(0)$
is obtained as an expression of $s(0)$ and the null data of $\psi_{ABCD}$  
of order $ \le n-2$.

For the quantities $\nabla_{\bf n}\Omega(0)$ the induction step follows immediately from 
(\ref{Omega-equ}) and for the quantities $\nabla_{\bf n}s(0)$ it follows with 
(\ref{s-equ}) by using (\ref{Omega-equ})  again.

\vspace{.2cm}

This proves the first part of the Lemma. The second statement follows because equation (\ref{psi=sym+rest}) shows that no restrictions 
are imposed by the field equations on the quantities 
$\nabla_{(E_1}\,^{(E'_1}\, \ldots \,\nabla_{E_n}\,^{E'_n)}\,\psi_{ABCD)}(0)$. By the argument given above all formal covariant derivatives  are given by algebraic expressions of the null data of $\psi$ at $p$ and these expression impose no restrictions on the null data. $\Box$

\vspace{.5cm}

By (\ref{normal-expansion}) the symmetric parts  of the covariant derivatives determined in Lemma \ref{unique-expansion} can be regarded as Taylor coefficients of corresponding tensor fields. By Borel's theorem (\cite{dieudonne:I})  we can then find smooth fields
$\hat{\Omega}$, $\hat{ \Pi }$, $\hat{\psi}_{ABCD}$, $\hat{\Phi}_{ABA'B'}$ near $p$  whose 
Taylor coefficients  at $p$ coincide with  the Taylor coefficients determined by the procedure above (but fairly arbitrary away from $p$). We can assume that these fields satisfy   near $p$ the symmetry and the reality properties discussed in section \ref{spinors}. With these fields we set 
$\hat{R}_{ABCC'DD'} = \hat{\Omega}\,\hat{\psi}_{ABCD}\,\epsilon_{C'D'}
+ \hat{\Phi}_{ABC'D'}\,\epsilon_{CD}$, which corresponds to the curvature spinor whose Taylor coefficients entered the discussion above, and define  
the `curvature tensor' 
$\hat{R}^i\,_{jkl}$ by following  (\ref{curvature-spin curvature}).

To decide whether these smooth fields do in fact satisfy the field equations at all orders at $p$ we first need to determine frame and connection coefficients consistent with the curvature tensor.

\section{The structural equations.}
\label{structural equations}

The frame and the connection coefficients which we want to
satisfy  the structural equations with the `curvature spinor' $\hat{R}_{ABCC'DD'}$
will be denoted in the following by  $\hat{e}^{\mu}\,_i $ and
$\hat{\Gamma}_{AA'}\,^{C}\,_{B}$. It turns out that  these functions are determined already by the subsystem 
\begin{equation}
\label{Xcontr-str-equ}
\hat{t}_k\,^i\,_l\,\hat{e}^{\mu}\,_i\,X^l = 0, \quad \quad (\hat{r}^A\,_{B\,kl} - \hat{R}^A\,_{B\,kl})\,X^k = 0,
\end{equation}
of the structural equations, 
where the fields 
$\hat{t}_k\,^i\,_l$ and $\hat{r}^A\,_{B\,kl}$ are given by the right hand sides of 
(\ref{torsion-tensor}), (\ref{torsion-tensor-r})
with $e$ and $\Gamma$ replaced by 
$\hat{e}$ and $\hat{\Gamma}$ and where  $X^i = \delta^i\,_{\mu}\,x^{\mu}$.
Assuming  (\ref{x-sigma}) and (\ref{nablaX-e=0}) to be satisfied by $\hat{e}^{\mu}\,_i $ and
$\hat{\Gamma}_{AA'}\,^{C}\,_{B}$, these equations can be written 
\begin{equation}
\label{Xcontr-torsion-free-cond}
\hat{e}^{\mu}\,_{k,\,\nu}\,x^{\nu} + \hat{e}^{\mu}\,_l\,(\delta^l_{\nu}\,\hat{e}^{\nu}\,_k - \delta^l_k)  
+ \hat{\Gamma}_k\,^i\,_l\,X^l\,\hat{e}^{\mu}\,_i = 0,
\end{equation}
\begin{equation}
\label{Xcontr-ricci-id}
\hat{\Gamma}_{l}\,^A\,_{B,\,\mu}\,x^{\mu}
+ \hat{\Gamma}_{k}\,^A\,_{B}\,\delta^k_{\mu}\,\hat{e}^{\mu}\,_{l}
+  \hat{\Gamma}_{l}\,^{j}\,_{k}\,X^k\,\hat{\Gamma}_{j}\,^A\,_B =  \hat{R}^A\,_{B\,kl}\,X^k,
\end{equation}
where the $ \hat{\Gamma}_k\,^i\,_l$ are given  in spinor notation by
\[
\hat{\Gamma}_{AA'}\,^{CC'}\,_{BB'} =  
\hat{\Gamma}_{AA'}\,^{C}\,_{B}\,\epsilon_{B'}\,^{C'}
+  \bar{\hat{\Gamma}}_{AA'}\,^{C'}\,_{B'}\,\epsilon_{B}\,^{C},
 \]
so that they are real and satisfy $ \hat{\Gamma}_{k\,i\,l} = -  \hat{\Gamma}_{k\,l\,i} $
as a consequence of  $\hat{\Gamma}_{l}\,_{AB} = \hat{\Gamma}_{l}\,_{(AB)}$. Equations 
(\ref{Xcontr-torsion-free-cond}), (\ref{Xcontr-ricci-id}) imply that a smooth solution $\hat{e}^{\mu}\,_i(x^{\mu})$,
$\hat{\Gamma}_{AA'}\,^{C}\,_{B}(x^{\mu})$ near $x^{\mu} = 0$ with  $\det(\hat{e}^{\mu}\,_i ) \neq 0$ must satisfy 
\begin{equation}
\label{e-Gamma-at-0}
\hat{e}^{\mu}\,_{k}(0) = \delta^{\mu}\,_k, \quad \quad  
\hat{\Gamma}_{l}\,^A\,_{B}(0) = 0.
\end{equation}

Equations (\ref{Xcontr-torsion-free-cond}), (\ref{Xcontr-ricci-id}) can be discussed by analysing the ODE's which are implied by them along the curves 
$\tau \rightarrow \tau\,x^{\mu}_*$, $x^{\mu}_* \neq 0$. These ODE's will be considered in section \ref{transport-equ-and-inner-constr}, for our present purpose a more direct approach will be sufficient. To simplify the algebra we rewrite the equations in terms of the unknowns 
\[
\hat{c}^{\mu}\,_{k} \equiv \hat{e}^{\mu}\,_{k} - \delta^{\mu}\,_k, \quad 
\hat{\Gamma}_{AA'}\,^{C}\,_{B},
\]
to obtain them in the form
\begin{equation}
\label{Ycontr-torsion-free-cond}
\hat{c}^{\mu}\,_{k,\,\nu}\,x^{\nu} 
+ \hat{c}^{\mu}\,_k  
+ \hat{c}^{\mu}\,_l\,\delta^l_{\nu}\,\hat{c}^{\nu}\,_k 
+ \hat{\Gamma}_k\,^i\,_l\,X^l\,\hat{c}^{\mu}\,_i 
+ \hat{\Gamma}_k\,^i\,_l\,X^l\,\delta^{\mu}\,_i 
= 0,
\end{equation}
\begin{equation}
\label{Ycontr-ricci-id}
\hat{\Gamma}_{l}\,^A\,_{B,\,\nu}\,x^{\nu}
+ \hat{\Gamma}_{l}\,^A\,_{B}
+ \hat{\Gamma}_{k}\,^A\,_{B}\,\delta^k_{\mu}\,\hat{c}^{\mu}\,_{l}
+  \hat{\Gamma}_{l}\,^{j}\,_{k}\,X^k\,\hat{\Gamma}_{j}\,^A\,_B -  \hat{R}^A\,_{B\,kl}\,X^k = 0.
\end{equation}
By taking formally partial derivatives, observing (\ref{e-Gamma-at-0}), and evaluating at $x^{\mu} = 0$ one obtains unique sequences of derivatives
\[
\hat{c}^{\mu}\,_{k,\,\nu_1\, \ldots \,\nu_k}(0), \quad  
\hat{\Gamma}_{l}\,^A\,_{B,\,\nu_1\, \ldots \,\nu_k}(0), \quad k \in \mathbb{N},
\]
which are symmetric in the indices $\nu_1\, \ldots \,\nu_k$ and are determined by
the partial derivatives of the field $\hat{R}^A\,_{B\,kl}$ at the origin. By Borel's theorem 
(\cite{dieudonne:I}) we can then find smooth fields
$\hat{c}^{\mu}\,_k$ and $\hat{\Gamma}_{l}\,^A\,_B$ near $x^{\mu} = 0$ whose Taylor coefficients coincide with the coefficients given above. Because of $ \hat{R}_{AB\,kl} =  \hat{R}_{(AB)\,kl}$ and the structure of the equations,
these fields can be chosen such that $\hat{c}^{\mu}\,_{k}$ is real and 
$\hat{\Gamma}_{l\,AB} = \hat{\Gamma}_{l\,(AB)}$. While the choice of the fields is rather arbitrary away from $x^{\mu} = 0$ they satisfy the structural equations at all orders at $x^{\mu}= 0$ so that  
\begin{equation}
\label{Zcontr-torsion-free-cond}
\hat{c}^{\mu}\,_{k,\,\nu}\,x^{\nu} 
+ \hat{c}^{\mu}\,_k  
+ \hat{c}^{\mu}\,_l\,\delta^l_{\nu}\,\hat{c}^{\nu}\,_k 
+ \hat{\Gamma}_k\,^i\,_l\,X^l\,\hat{c}^{\mu}\,_i 
+ \hat{\Gamma}_k\,^i\,_l\,X^l\,\delta^{\mu}\,_i 
= O(|x|^{\infty}),
\end{equation}
\begin{equation}
\label{Zcontr-ricci-id}
\hat{\Gamma}_{l}\,^A\,_{B,\,\nu}\,x^{\nu}
+ \hat{\Gamma}_{l}\,^A\,_{B}
+ \hat{\Gamma}_{k}\,^A\,_{B}\,\delta^k_{\mu}\,\hat{c}^{\mu}\,_{l}
+  \hat{\Gamma}_{l}\,^{j}\,_{k}\,X^k\,\hat{\Gamma}_{j}\,^A\,_B -  \hat{R}^A\,_{B\,kl}\,X^k 
= O(|x|^{\infty}),
\end{equation}
where the symbols $ O(|x|^{\infty})$ on the right hand sides  indicate that the quantities on the left hand side are  for all $n \in \mathbb{N}$ of the order $ O(|x|^n)$ as $x^{\mu} \rightarrow 0$.

With (\ref{e-Gamma-at-0}) it follows that 
\begin{equation}
\label{Bc-at-0}
\hat{c}^{\mu}\,_k(0) = 0, \quad \hat{c}^{\mu}\,_{k, \nu}(0) = 0, \quad  
\hat{\Gamma}_{l}\,^A\,_{B}(0) = 0.
\end{equation}
We restrict the following discussion to  some neighbourhood of the origin on which 
the smooth field $\hat{e}^{\mu}\,_k \equiv \delta^{\mu}\,_k + \hat{c}^{\mu}\,_k$ satisfies 
$\det(\hat{e}^{\mu}\,_k) \neq 0$. It is there orthonormal for the metric $\hat{g}_{\mu \nu} \equiv \eta_{ij}\,\hat{\sigma}^i\,_{\mu}\,\hat{\sigma}^j\,_{\nu}$, where the 
$\hat{\sigma}^i\,_{\mu}$ denote the 1-forms dual to the $\hat{e}^{\mu}\,_{k}$. 
Because $\hat{\Gamma}_{i\,AB} = \hat{\Gamma}_{i\,BA}$,
whence $\hat{\Gamma}_{i\,j\,k} = - \hat{\Gamma}_{i\,k\,j}$, the connection $\hat{\nabla}$ defined by $\hat{e}^{\mu}\,_{k}$ and 
$\hat{\Gamma}_i\,^j\,_k$ resp. $\hat{\Gamma}_i\,^A\,_B$, which satisfies  for instance 
$\hat{\nabla}_i\,\hat{e}_k = \hat{\Gamma}_i\,^j\,_k\,\hat{e}_k$ with $\hat{\nabla}_i \equiv \hat{\nabla}_{\hat{e}_i}$, is $\hat{g}$-metric compatible  in the sense that $\hat{\nabla}\,\hat{g} = 0$. 

The symmetries of the fields $\hat{\Gamma}_{k\,AB}$ and $\hat{R}_{AB\,jk}$ imply the following results.

\vspace{.2cm}

\begin{lemma}
\label{x-ghat-normal}
(i) The coordinates $x^{\mu}$  and the frame coefficients $\hat{e}^{\mu}\,_k$ satisfy the requirements  (\ref{x-sigma}), (\ref{sigma-x}), (\ref{nablaX-e=0})
of a normal gauge at all orders at $x^{\mu}  = 0$, so that 
\begin{equation}
\label{Zx-sigma}
(\hat{e}^{\nu}\,_j(x) - \delta^{\nu}\,_j)\,\delta^j_{\mu}\,x^{\mu} = O(|x|^{\infty}),  
\end{equation}
\begin{equation}
\label{Zsigma-x}
x^{\mu}\,\eta_{\mu \nu}\,(\hat{e}^{\nu}\,_k(x) - \delta^{\nu}\,_k) = O(|x|^{\infty}),
\end{equation}
\begin{equation}
\label{ZnablaX-e=0}
 \delta^k_{\nu}\,x^{\nu}\,\hat{\Gamma}_k\,^i\,_j(x) =  O(|x|^{\infty}).
\end{equation}
(ii) Consider the curve 
$\tau \rightarrow x^{\mu}(\tau) = \tau\,x^{\mu}_*$, $x^{\mu}_* \neq 0$, through the origin.
The components of its tangent vectors $\dot{x}^{\mu} = x_*^{\mu}$ in the frame 
$\hat{e}^{\mu}\,_k$, given by $z^k(\tau) = x^{\mu}_*\,\hat{\sigma}^k\,_{\mu}(\tau\,x_*)$,
satisfies 
\begin{equation}
\label{tang-vec-approx}
z^k(\tau) - \delta^k\,_{\mu}\,x_*^{\mu} = O(|\tau\,x_*|^{\infty}),
\end{equation}
the curve satisfies the geodesic equation at all orders at $\tau = 0$, 
\begin{equation}
\label{geod-approx}
\hat{\nabla}_{\dot{x}} \,\dot{x} = O(|\tau\,x_*|^{\infty}),
\end{equation}
and the frame 
$\hat{e}_{k} = \hat{e}^{\mu}\,_{k}\,\partial_{x^{\mu}}$ satisfies the equation of parallel transport along these curves at all orders at $\tau = 0$, 
\begin{equation}
\label{frame-par-transp--approx}
\hat{\nabla}_{\dot{x}} \,\hat{e}_k = O(|\tau\,x_*|^{\infty}).
\end{equation}
\end{lemma}

\vspace{.2cm}

\noindent
{\bf Proof}: To obtain the relations  (\ref{Zx-sigma}), (\ref{Zsigma-x}), (\ref{ZnablaX-e=0}) we 
contract (\ref{Zcontr-torsion-free-cond}) and (\ref{Zcontr-ricci-id})   with 
$\delta^k_{\mu}\,x^{\mu}$ and $\delta^l_{\mu}\,x^{\mu}$ respectively to obtain the relations
\begin{equation}
\label{ZZcontr-torsion-free-cond}
\hat{c}^{\mu}\,_{,\,\nu}\,x^{\nu} 
+  \hat{c}^{\mu}\,_l\,\delta^l_{\nu}\,\hat{c}^{\nu}
+ \hat{\Gamma}^i\,_l\,X^l\,(\hat{c}^{\mu}\,_i  + \delta^{\mu}\,_i) 
= O(|x|^{\infty}),
\end{equation}
\begin{equation}
\label{ZZcontr-ricci-id}
\hat{\Gamma}^A\,_{B,\,\nu}\,x^{\nu}
+ \hat{\Gamma}_{k}\,^A\,_{B}\,\delta^k_{\mu}\,\hat{c}^{\mu}\,
+  \hat{\Gamma}^{j}\,_{k}\,X^k\,\hat{\Gamma}_{j}\,^A\,_B  
= O(|x|^{\infty}).
\end{equation}
for the quantities
\[
\hat{c}^{\nu} \equiv \hat{c}^{\nu}\,_j(x)\,\delta^j_{\mu}\,x^{\mu},  \,\,\,
\hat{c}^{\nu}\,_k \equiv x^{\mu}\,\eta_{\mu \nu}\,\hat{c}^{\nu}\,_k(x), \,\,\,
\hat{\Gamma}^A\,_B \equiv  \delta^k_{\nu}\,x^{\nu}\,\hat{\Gamma}_k\,^A\,_B(x), \,\,\,
\hat{\Gamma}^i\,_j \equiv  \delta^k_{\nu}\,x^{\nu}\,\hat{\Gamma}_k\,^i\,_j(x).
\]
If $\hat{c}^{\mu} = O(|x|^p)$ and $\hat{\Gamma}^A\,_B = O(|x|^q)$  with some 
$p, q \in \mathbb{N}$, these relations imply with (\ref{Bc-at-0}) relations of the form 
\[
\hat{c}^{\mu}\,_{,\,\nu}\,x^{\nu} = O(|x|^{p+2}) + O(|x|^{q + 1}), \quad
\hat{\Gamma}^A\,_{B,\,\nu}\,x^{\nu} = O(|x|^{p+1}) + O(|x|^{q + 2}).
\]
Because $\hat{c}^{\mu} = O(|x|^3)$ and $\hat{\Gamma}^A\,_B = O(|x|^2)$
by  (\ref{Bc-at-0}), the second relation implies that 
$\hat{\Gamma}^A\,_{B,\,\nu}\,x^{\nu} = O(|x|^4)$ whence also 
$\hat{\Gamma}^A\,_{B} = O(|x|^4)$
and the first relation gives then 
$\hat{c}^{\mu}\,_{,\,\nu}\,x^{\nu} = O(|x|^{5})$
whence $\hat{c}^{\mu} = O(|x|^{5})$. Repeating the argument we conclude that 
$\hat{c}^{\mu} = O(|x|^{\infty})$ and $\hat{\Gamma}^A\,_B = O(|x|^{\infty})$,
which are the relations
(\ref{Zx-sigma}) and (\ref{ZnablaX-e=0}).
 
Observing  that
\[
\hat{\Gamma}_k\,^i\,_l\,X^l\,\delta^{\mu}\,_i \,x^{\lambda}\,\eta_{\lambda \mu} =
\hat{\Gamma}_{k\,j\,l}\,X^l\,\eta^{ji}\,\delta^{\mu}\,_i \,x^{\lambda}\,\eta_{\lambda \mu} =
\hat{\Gamma}_{k\,j\,l}\,X^j\,X^l = 0,
\]
the contraction of  (\ref{Zcontr-torsion-free-cond}) with $x^{\lambda}\,\eta_{\lambda \mu}$
gives
\begin{equation}
\label{2ZZcontr-torsion-free-cond}
\hat{c}_{k,\,\nu}\,x^{\nu}   
+ \hat{c}_l\,\delta^l_{\nu}\,\hat{c}^{\nu}\,_k 
+ \hat{\Gamma}_k\,^i\,_l\,X^l\,\hat{c}_i 
+ \hat{\Gamma}_k\,^i\,_l\,X^l\,\delta^{\mu}\,_i 
= O(|x|^{\infty}),
\end{equation}
which implies with the previous result that 
$\hat{c}_k  = O(|x|^{\infty})$, which is in fact (\ref{Zsigma-x}).

\vspace{.2cm}

Contraction of the relation
$(\hat{e}^{\nu}\,_j(\tau\,x_*) - \delta^{\nu}\,_j)\,\delta^j_{\mu}\,x^{\mu}_* = O(|\tau\,x_*|^{\infty})$,
which holds  by  (\ref{Zx-sigma}), with $- \hat{\sigma}^k\,_{\nu}(\tau\,x_*)$ gives 
(\ref{tang-vec-approx}). In terms of the frame one has
\[
(\hat{\nabla}_{\dot{x}} \,\dot{x})^k = \frac{d}{d\tau}z^k 
+ z^j\,\hat{\Gamma}_j\,^k\,_l(\tau\,x_*)\,z^l = O(|\tau\,x_*|^{\infty}),
\]
and
\[
\hat{\nabla}_{\dot{x}} \,\hat{e}_k = z^j\,\,\hat{\Gamma}_j\,^l\,_k(\tau\,x_*)\,\hat{e}_l = 
O(|\tau\,x_*|^{\infty}),
\]
by (\ref{tang-vec-approx}) and (\ref{ZnablaX-e=0}).
$\Box$

\section{Formal and factual derivatives}
\label{f-and-f-derivatives}

The subsystem  (\ref{Xcontr-str-equ}) of the structural equations determines the functions 
$\hat{e}^{\mu}\,_{k}$  and $\hat{\Gamma}_i\,^j\,_k$ uniquely and implies that 
\begin{equation}
\label{t-free-curv-equ}
\hat{t}_i\,^j\,_k = O(|x|^n), \quad \quad \hat{r}^h\,_{k j l} -  \hat{R}^h\,_{k j l} = O(|x|^n)
\end{equation}
with $n = 1$. Moreover, direct calculations involving (\ref{e-Gamma-at-0}),  
(\ref{psi-0-1-data}), (\ref{Phi-0-1-data}), (\ref{Omega-0 - - 5-data}), (\ref{s-0 -- 3 - data}) 
show that  
\begin{equation}
\label{scal-form=act}
\hat{\nabla}_{\bf k}\,\hat{\Omega}(0) = \nabla_{\bf k}\,\Omega(0), \quad
\hat{\nabla}_{\bf k}\,\hat{ \Pi }(0) = \nabla_{\bf k}\, \Pi (0), 
\end{equation}
for $|{\bf k}| \le 3$ and 
\begin{equation}
\label{ten-form=act}
\hat{\nabla}_{\bf k}\hat{\Phi}_{ABA'B'}(0) = \nabla_{\bf k}\Phi_{ABA'B'}(0), \quad
\hat{\nabla}_{\bf k}\hat{\psi}_{ABCD}(0) = \nabla_{\bf k}\psi_{ABCD}(0),
\end{equation}
whence 
\begin{equation}
\label{curv-funct=form-curv}
\hat{\nabla}_{\bf k}\hat{R}^i\,_{jkl}(0) = \nabla_{\bf k}R^i\,_{jkl}(0), 
\end{equation}
for $|{\bf k}| \le 1$
where on the right hand sides are given the formal expressions derived in the previous section and on the left hand sides the factual covariant derivatives of the smooth fields 
$\hat{\Omega}$,  $\hat{s}$, $\hat{\Phi}_{ABA'B'}$, 
$\hat{\psi}_{ABCD}$, $\hat{R}^i\,_{jkl}$ at the point $x^{\mu} = 0$ with respect to the connection $\hat{\nabla}$. 
These relations imply that 
\begin{equation}
\label{Omega-hat-equ-O-n} 
\hat{\nabla}_{AA'}\,\hat{\nabla}_{BB'}\,\hat{\Omega} + \hat{\Omega}\,\hat{\Phi}_{ABA'B'} 
- \hat{ \Pi }\,\epsilon_{AB}\,\epsilon_{A'B'} = O(|x|^n),  
\end{equation}
\begin{equation}
\label{s-hat-equ-O-n}
\hat{\nabla}_{AA'}\,\hat{ \Pi } +\hat{ \nabla}^{BB'}\hat{\Omega}\,\hat{\Phi}_{ABA'B'} = O(|x|^n), 
\end{equation}
\begin{equation}
\label{Phi-hat-equ-O-n}
\hat{\nabla}_F\,^{D'}\hat{\Phi}_{ABF'D'} 
-  \hat{\nabla}^{C}\,_{F'}\hat{\Omega}\,\,\hat{\psi}_{ABCF}
= O(|x|^n),
\end{equation}
\begin{equation}
\label{pii-hat-equ-O-n} 
\hat{\nabla}^{C}\,_{F'}\hat{\psi}_{ABCF} = O(|x|^n),  
\end{equation}
hold with $n = 1$. Because the quantities $ \nabla_{\bf k}R^i\,_{jkl}(0)$ have been determined by invoking  the Bianchi identities  (see the discussion of  (\ref{Bianchi-2a}), (\ref{Bianchi-2b}))
it follows from (\ref{curv-funct=form-curv}) that 
\begin{equation}
\label{R-hat-Bainchi-O-n}
\sum_{cycl(ijl)}\hat{\nabla}_i\,\hat{R}^h\,_{k j l} = O(|x|^n),
\end{equation}
with $n = 1$. The purpose of this section is to derive the following result. 

\begin{proposition}
\label{main-result}
Relations (\ref{t-free-curv-equ}) to 
(\ref{R-hat-Bainchi-O-n}) hold true for all integers 
$n \in \mathbb{N}$ resp. multi-indices ${\bf k}$.
\end{proposition}

\begin{remark}
The following argument covers in particular the vacuum case in which 
$\Omega$ is set equal to $1$   and the only non-trivial fields are given by $e^{\mu}\,_k$, $\Gamma_i\,^j\,_k$ and $\psi_{ABCD}$.
\end{remark}

Before we begin with the proof we need to make a few observations. Because only a subsystem of the structural equations has been used so far, it is  not clear whether  the order relations (\ref{t-free-curv-equ}) hold for all $n \in \mathbb{N}$. The following result shows in particular how this question is related 
to the Bianchi identity (\ref{R-hat-Bainchi-O-n}).

 \begin{lemma}
 \label{t-hat-r-Rhat-control}
Denote by $\hat{\gamma}_k\,^i\,_l$ the connection coefficients  of the Levi-Civita connection of the metric $\hat{g}_{\mu \nu} = \eta_{jk}\,\hat{\sigma}^j\,_{\mu}\,\hat{\sigma}^k\,_{\nu}$ with respect to the frame $\hat{e}_k$. If the torsion tensor $\hat{t}_i\,^j\,_k$ of the connection 
$\hat{\nabla}$ behaves as $\hat{t}_i\,^j\,_k =  O(|x|^N)$ for some $N \in \mathbb{N}$, $N \ge 1$, then $\hat{\Gamma}_k\,^i\,_l - \hat{\gamma}_k\,^i\,_l =  O(|x|^N)$.
 
If $N \in \mathbb{N}$, $N \ge 1$, and
\begin{equation}
\label{hat-2nd-Bianchi-N}
\sum_{cycl(ijl)}\hat{\nabla}_i\,\hat{R}^h\,_{k j l} = O(|x|^N),
\end{equation}
 then 
$\,\,\hat{t}_j\,^k\,_l = O(|x|^{N+2})$,   
$\,\,\hat{r}^h\,_{k j l} - \hat{R}^h\,_{k j l}  = O(|x|^{N+1})$.

 \end{lemma}

\noindent
{\bf Proof}: Denote by $\hat{c}_l\,^j\,_k$ the commutator coefficients satisfying 
$[\hat{e}_l\,\hat{e}_k] = \hat{c}_l\,^j\,_k\,\hat{e}_j$. 
With $\hat{c}_{l\,i\,k} = \hat{c}_l\,^j\,_k\,\eta_{ji}$ and 
 $\hat{t}_{l\,i\,k} = \hat{t}_l\,^j\,_k\,\eta_{ji}$ the torsion free relation can be written
$\hat{\Gamma}_{l\,i\,k} - \hat{\Gamma}_{k\,i\,l} - \hat{c}_{l\,i\,k} 
= \hat{t}_{k\,i\,l}$.
It is well known that this implies
\[
2\,\hat{\Gamma}_{kli} - \{\hat{c}_{l\,i\,k} + \hat{c}_{k\,l\,i}  - \hat{c}_{i\,k\,l}\}
= \hat{t}_{l\,i\,k} + \hat{t}_{k\,l\,i} - \hat{t}_{i\,k\,l}.
\]
The same relations hold with $ \hat{t}_{l\,i\,k} = 0$ if
$\hat{\Gamma}_{l\,i\,k}$ is replaced by $\hat{\gamma}_{l\,i\,k}$. This gives 
\[
2\,(\hat{\Gamma}_{kli} - \hat{\gamma}_{kli})
= \hat{t}_{l\,i\,k} + \hat{t}_{k\,l\,i} - \hat{t}_{i\,k\,l},
\]
which implies the desired result.

The connection $\hat{\nabla}$ defined by 
$\hat{e}_{k}$ and $\hat{\Gamma}_i\,^j\,_k$ is metric compatible but at this stage not known to be 
torsion free. As pointed out in section \ref{conformal field equations}, the Bianchi identities
for the torsion tensor $\hat{t}_i\,^j\,_k$ and the curvature tensor $\hat{r}^i\,_{jkl}$
then take the form
\begin{equation}
\label{hat-1st-Bianchi}
\sum_{cycl(ijl)}\hat{\nabla}_i\,\hat{t}_j\,^k\,_l =
\sum_{cycl(ijl)}(\hat{r}^k\,_{ijl} - \hat{t}_i\,^m\,_j\,\hat{t}\,_m\,^k\,_l),
\end{equation}
\begin{equation}
\label{hat-2nd-Bianchi}
\sum_{cycl(ijl)}\hat{\nabla}_i\,\hat{r}^h\,_{k j l} =
\sum_{cycl(ijl)}\hat{t}_j\,^m\,_i\,\hat{r}^h\,_{kml}.
\end{equation}
 By the symmetries and reality conditions of the fields defining $\hat{R}^k\,_{ijl} $ the arguments which led to (\ref{Bianchi-1}) imply $\sum_{cycl(ijl)}\,\hat{R}^k\,_{ijl} = 0$ near $p$.
 Equation (\ref{hat-1st-Bianchi}) can thus be written
\[
\sum_{cycl(ijl)}\hat{\nabla}_i\,\hat{t}_j\,^k\,_l =
\sum_{cycl(ijl)}(\hat{r}^k\,_{ijl} - \hat{R}^k\,_{ijl}  - \hat{t}_i\,^m\,_j\,\hat{t}\,_m\,^k\,_l).
\]
Transvecting this equation with $X^i$, observing (\ref{Xcontr-str-equ}) and the anti-symmetry of the torsion tensor gives
\[
X^i\,\hat{\nabla}_i\,\hat{t}_j\,^k\,_l + \hat{t}\,_j\,^k\,_i\,\hat{\nabla}_l X^i 
+ \hat{\nabla}_j X^i \,\hat{t}\,_i\,^k\,_l =
x^i\,(\hat{r}^k\,_{ijl} - \hat{R}^k\,_{ijl} ).
\]
Similarly, transvecting the rewrite 
\[
\sum_{cycl(ijl)}\hat{\nabla}_i\,(\hat{r}^h\,_{k j l} - \hat{R}^h\,_{k j l}) =
\sum_{cycl(ijl)}\hat{t}_j\,^m\,_i\,\hat{r}^h\,_{kml} 
- \sum_{cycl(ijl)}\hat{\nabla}_i\,\hat{R}^h\,_{k j l},
\]
of (\ref{hat-2nd-Bianchi}) with $X^i$ gives 
\[
X^i\,\hat{\nabla}_i\,(\hat{r}^h\,_{k j l} - \hat{R}^h\,_{k j l}) 
+     (\hat{r}^h\,_{k j i} - \hat{R}^h\,_{k j i})\,\hat{\nabla}_l X^i 
+      \hat{\nabla}_j X^i \,(\hat{r}^h\,_{k i l} - \hat{R}^h\,_{k i l})
\]
\[
=   \hat{t}_l\,^m\,_j\,\hat{r}^h\,_{kmi}\,X^i 
- X^i\left(\sum_{cycl(ijl)}\hat{\nabla}_i\,\hat{R}^h\,_{k j l}\right).
\]
The result follows now with  (\ref{X-definition}) by taking derivatives and evaluating at 
$x^{\mu} = 0$. $\hfill \Box$

\vspace{.5cm}

Assume that there exist a smooth solution to the field equations in the given gauge which induces the prescribed null data at $p$. By the arguments given above the  $\infty$-jet of the solution at $p$ must then coincide with the
expressions on the right hand sides of (\ref{scal-form=act}), (\ref{ten-form=act}), (\ref{curv-funct=form-curv}). It is not obvious, however, that it must also coincide with the $\infty$-jets of the functions $\hat{e}^{\mu}\,_k$, $\hat{\Gamma}_i\,^j\,_k$, $\hat{\Omega}$,  $\hat{ \Pi }$, $\hat{\Phi}_{ABA'B'}$, $\hat{\psi}_{ABCD}$ at $p$.
The reason is, that, following  (\ref{normal-expansion}),  these functions have been defined
so that their Taylor coefficients at $x^{\mu} = 0$ coincide with the {\it symmetrized} derivatives $\nabla_{(\bf k)}\,\Omega(0)$, $\nabla_{(\bf k)}\, \Pi (0)$, 
$\nabla_{(\bf k)}\Phi_{ABA'B'}(0)$, $\nabla_{(\bf k)}\psi_{ABCD}(0)$ and it  is not clear how much of the information encoded in  the  
unsymmetrized derivatives is transported by the symmetrized derivatives. 
In particular,
while the Bianchi identities are by 
(\ref{Bianchi-2a}), (\ref{Bianchi-2b}) part of  the conformal field equations
and the coefficients on the right hand sides of  (\ref{scal-form=act}), (\ref{ten-form=act}), (\ref{curv-funct=form-curv}) have been determined so as to satisfy these identities,  it is not obvious at this stage 
that relation (\ref{hat-2nd-Bianchi-N}) should be satisfied for  integers $N > 1$.

\vspace{.5cm}

\noindent
{\bf Proof of Proposition \ref{main-result}}: The induction argument to be given below  will make use of  
the following general considerations. 
Let $T_{A_1 \ldots A_j\,B'_1 \ldots B'_k}$ denote a smooth spinor field and $\nabla$ a metric compatible connection with curvature tensor $r^i\,_{jkl}$ and torsion tensor $t_i\,^j\,_k$. To begin with assume that $t_i\,^j\,_k = 0$.
If the derivatives on the right hand side of the symmetrization formula 
\[
\nabla_{(i_1}  \ldots  \nabla_{i_n)}T_{A_1 \ldots A_j\,B'_1 \ldots B'_k}
= \frac{1}{n!} \sum_{\pi \in {\cal S}_n} 
\nabla_{i_{\pi(1)}}  \ldots  \nabla_{i_{\pi(n)}}T_{A_1 \ldots A_j\,B'_1 \ldots B'_k}
\]
are then commuted to bring them into their natural order, one obtains an equation of the form
\[
\nabla_{\bf n} \,T_{A_1 \ldots A_j\,B'_1 \ldots B'_k} =
\nabla_{(\bf n)}\,T_{A_1 \ldots A_j\,B'_1 \ldots B'_k} + 
C^*_{{\bf n}\,\,A_1 \ldots A_j\,B'_1 \ldots B'_k},
\]
where the spinor field $C^*$ is a sum of terms which depend on the covariant derivatives 
of $T$ and $r^i\,_{jkl}$  of order $\le  {|\bf n|} - 2$.
Using these formulas to substitute successively  in the formulas for $n = 3, 4, \ldots$ 
the covariant derivatives of $T$ of lower order by their symmetric parts one obtains  formulas 
\begin{equation}
\label{sym-nonsym-cov-der}
\nabla_{\bf n} \,T_{A_1 \ldots A_j\,B'_1 \ldots B'_k} =
\nabla_{(\bf n)}\,T_{A_1 \ldots A_j\,B'_1 \ldots B'_k} + 
C_{{\bf n}\,\,A_1 \ldots A_j\,B'_1 \ldots B'_k}, \quad |{\bf n}| \ge 0,
\end{equation}
with spinor valued functions
\[
C_{\bf n} =
C_{\bf n}(\nabla_{(\bf p)}T,  \nabla_{\bf q}\,r)
\quad \mbox{where} \quad 
|{\bf p}|, |{\bf q}| \le  {|\bf n|} - 2,
\]
which satisfy $C_{(\bf n)} = 0$.
These formulas show how the covariant derivatives of $T$ at the point $x = 0$ are determined from the Taylor coefficients in (\ref{normal-expansion}) and the derivatives 
of  the curvature tensor 
at $x = 0$. 

\vspace{.2cm}

Formulas (\ref{sym-nonsym-cov-der}) represent universal relations. The functions  
$C_{\bf n}$  depend on the connection $\nabla$  only via the derivatives $\nabla_{\bf q}\,r$ of its curvature tensor. (We ignore the fact that the explicit dependence of $C_{\bf n}$ on the $\nabla_{\bf q}\,r$ may be written in different forms by using the symmetries and the differential identities satisfied by the curvature tensor). The full index notation of 
(\ref{sym-nonsym-cov-der}) emphasizes that the explicit  structure of the  functions $C_{\bf n}$ does depend on the index type of the spinor field $T$ and in following equations we shall write out the appropriate indices.

\vspace{.2cm}

With the notation of Section \ref{formal expansion} the unknowns in the field equations must have representations of the form
\begin{equation}
\label{start-Omega-n-ord-sym-decomp}
\nabla_{\bf n} \,\Omega =
\nabla_{(\bf n)}\,\Omega + 
C_{{\bf n}}(\nabla_{(\bf p)} \,\Omega, \nabla_{\bf q} \,R),
\end{equation}
\begin{equation}
\label{start-s-n-ord-sym-decomp}
\nabla_{\bf n} \, \Pi  =
\nabla_{(\bf n)}\, \Pi  + 
C_{{\bf n}}(\nabla_{(\bf p)} \,\Pi, \nabla_{\bf q} \,R),\,\,
\end{equation}
\begin{equation}
\label{start-Phi-n-ord-sym-decomp}
\nabla_{\bf n} \,\Phi_{ABA'B'} =
\nabla_{(\bf n)}\,\Phi_{ABA'B'} + 
C_{{\bf n}\,\,ABA'B'}(\nabla_{(\bf p)} \,\Phi, \nabla_{\bf q} \,R),
\end{equation}
\begin{equation}
\label{start-psi-n-ord-sym-decomp}
\nabla_{\bf n} \,\psi_{ABCD} =
\nabla_{(\bf n)}\,\psi_{ABCD} + 
C_{{\bf n}\,\,ABCD}(\nabla_{(\bf p)} \,\psi, \nabla_{\bf q} \,R),
\end{equation}
with $|{\bf p}|, |{\bf q}| \le  {|\bf n|} - 2$ and quantities
$ \nabla_{\bf q} \,R$ which are understood as derivatives of the curvature defined by $\nabla$.
Using these as a starting point we can impose equations 
(\ref{Omega-equ}) to (\ref{gauged-spin-curvature-decomposition}) and 
proceed as in  Section \ref{formal expansion}  to derive for all multi-indices ${\bf k}$ expressions  for the quantities
$\nabla_{\bf k}\,\Omega(0)$, 
$\nabla_{\bf k}\, \Pi (0)$, 
$\nabla_{\bf k}\Phi_{ABA'B'}(0)$, 
$\nabla_{\bf k}\psi_{ABCD}(0)$,
and thus also for 
$ \nabla_{\bf k}R^i\,_{jkl}(0)$
in terms of the null data, which  
are given by the totally symmetric part of $\hat{\nabla}_{\bf n} \,\hat{\psi}_{ABCD}(0)$. 
These expressions could be inserted into the equations above but for the sake of comparison it will be better not to 
do this here.

Formulas  (\ref{sym-nonsym-cov-der}) do not immediately apply to the functions 
$\hat{\Omega}$,  $\hat{ \Pi }$, $\hat{\Phi}_{ABA'B'}$, $\hat{\psi}_{ABCD}$ with the connection 
$\hat{\nabla}$ and the curvature tensor $\hat{r}^i\,_{jkl}$.
They can be generalized, however,  to the case where
the connection $\nabla$ is not torsion free by observing (\ref{gen-comm}).
The functions $C_{\bf n}$ will then depend on the symmetrized derivatives 
$\nabla_{(\bf k)}T_{A_1 \ldots A_j\,B'_1 \ldots B'_k}$ of order  ${|\bf k|} \le {|\bf n|} - 1$ and  on the derivatives  of the torsion as well as on those of the curvature tensor.

Consider now the point $p$ with $x^{\mu}(p) = 0$ and assume that 
$t_i\,^j\,_k(x^{\mu}) = O(|x|^{N'})$ with some integer $N' \ge 1$ as $x^{\mu} \rightarrow 0$. 
It follows then with (\ref{gen-comm}) that the restriction of  (\ref{sym-nonsym-cov-der}) to the point $x^{\mu} = 0$ 
is valid as it stands if $|{\bf n}| \le N' + 1$. {\it At that point } we thus get 
for $|{\bf n}| \le N' + 1$ the relations 

\begin{equation}
\label{hat-Omega-n-ord-sym-decomp}
\hat{\nabla}_{\bf n} \,\hat{\Omega} =
\hat{\nabla}_{(\bf n)}\,\hat{\Omega} + 
C_{{\bf n}}(\hat{\nabla}_{(\bf p)} \,\hat{\Omega}, \hat{\nabla}_{\bf q} \,\hat{r}),
\end{equation}
\begin{equation}
\label{hat-s-n-ord-sym-decomp}
\hat{\nabla}_{\bf n} \,\hat{ \Pi } =
\hat{\nabla}_{(\bf n)}\,\hat{ \Pi } + 
C_{{\bf n}}(\hat{\nabla}_{(\bf p)} \,\hat{\Pi}, \hat{\nabla}_{\bf q} \,\hat{r}),\,\,
\end{equation}
\begin{equation}
\label{hat-Phi-n-ord-sym-decomp}
\hat{\nabla}_{\bf n} \,\hat{\Phi}_{ABA'B'} =
\hat{\nabla}_{(\bf n)}\,\hat{\Phi}_{ABA'B'} + 
C_{{\bf n}\,\,ABA'B'}(\hat{\nabla}_{(\bf p)} \,\hat{\Phi}, \nabla_{\bf q} \,\hat{r}),
\end{equation}
\begin{equation}
\label{hat-psi-n-ord-sym-decomp}
\hat{\nabla}_{\bf n} \,\hat{\psi}_{ABCD} =
\hat{\nabla}_{(\bf n)}\,\hat{\psi}_{ABCD} + 
C_{{\bf n}\,\,ABCD}(\hat{\nabla}_{(\bf p)} \,\hat{\psi}, \hat{\nabla}_{\bf q} \,\hat{r}),
\end{equation}
with $|{\bf p}|, |{\bf q}| \le  {|\bf n|} - 2$ and functions $C_{\bf n}$ which are identical with those appearing in the corresponding equation in (\ref{start-Omega-n-ord-sym-decomp}) to 
(\ref{start-psi-n-ord-sym-decomp}).

To compare these two sets of equations we observe that  only the properties (\ref{x-sigma}) and (\ref{nablaX-e=0}) of the frame and the connection coefficients have been used
to derive the normal expansion (\ref{normal-expansion}).
Because  these are satisfied by Lemma \ref{x-ghat-normal} also by the coefficients 
$\hat{e}^{\mu}\,_{k}$ and $ \hat{\Gamma}_i\,^j\,_k$, the normal expansions
of the fields $\hat{\Omega}$,  $\hat{s}$, $\hat{\Phi}_{ABA'B'}$, $\hat{\psi}_{ABCD}$ can thus be expressed  in terms of the derivatives with respect to the connection $\hat{\nabla}$. This implies   
$\hat{\nabla}_{(\bf k)}\hat{\Omega} = \nabla_{(\bf k)}\Omega$, 
$\hat{\nabla}_{(\bf k)}\hat{ \Pi } = \nabla_{(\bf k)} \Pi $, 
$\hat{\nabla}_{(\bf k)}\hat{\Phi}_{ABA'B'} = \nabla_{(\bf k)}\Phi_{ABA'B'}$, 
$\hat{\nabla}_{(\bf k)}\hat{\psi}_{ABCD} = \nabla_{(\bf k)}\psi_{ABCD}$
for all multi-indices ${\bf k}$ 
(here and in the following all spinors are thought  to be taken at the point $x^{\mu} = 0$).
It follows that the right hand sides of the two sets of equations are distinguished now only by the occurrence of the spinors 
$\nabla_{\bf q} \,R$ in the first set and 
the spinors $\hat{\nabla}_{\bf q} \,\hat{r}$ in the second set. 

Consider now as a induction hypothesis
the relations (\ref{scal-form=act}), (\ref{ten-form=act})  (\ref{curv-funct=form-curv}) with multi-indices ${\bf k}$ such that  $|{\bf k}| \le N'$.  Because the formal derivatives of the tensor $R^i\,_{jkl}$ have 
been determined such that the Bianchi identities are satisfied at all orders, relations 
 (\ref{curv-funct=form-curv}) imply that (\ref{hat-2nd-Bianchi-N}) holds with $N = N'$. 
It follows then from  Lemma \ref{t-hat-r-Rhat-control} that the assumption above on the torsion tensor is  satisfied and  (\ref{curv-funct=form-curv}) implies with the Lemma that 
$\hat{\nabla}_{\bf q}\hat{r}^h\,_{k j l} = \hat{\nabla}_{\bf q}\hat{R}^h\,_{k j l}
=  \nabla_{\bf q}R^h\,_{k j l}$ with $|{\bf q}| \le N'$.
Comparing the two sets of equations above we can obtain relations
 (\ref{scal-form=act}), (\ref{ten-form=act})  (\ref{curv-funct=form-curv}) with multi-indices 
 ${\bf k}$ such that  $|{\bf k}| = N' + 1$.  

With the properties noted in the beginning of this section this implies that  (\ref{scal-form=act}), (\ref{ten-form=act})  (\ref{curv-funct=form-curv}) hold true for  multi-indices ${\bf k}$ of all orders. It follows that the order relations 
(\ref{t-free-curv-equ}) and  (\ref{Omega-hat-equ-O-n}) to (\ref{R-hat-Bainchi-O-n}) are true for all integers $n \in \mathbb{N}$. $\Box$

\section{Transport equations and inner constraints.}
\label{transport-equ-and-inner-constr}

We have prescribed the radiation field, read off the null data at the vertex $p$, and constructed sequences of expansion coefficients at $p$ which can be realized as $\infty$-jets at $p$ of smooth fields which satisfy the (conformal) field equations at all orders at $p$. We want to discuss now which information can be derived from the radiation field  in some neighbourhood of $p$ on ${\cal N}_p$.

By definition, the characteristics of any hyperbolic system of first order are  those hypersurfaces on which  the system induces inner equations on (combinations of)  the dependent variables. 
On the other hand, the (conformal) Einstein equations induce as a consequence of  their gauge freedom constraints on their Cauchy data on any hypersurface.
On null hypersurfaces, which represent the characteristics of the  (conformal) Einstein equations, these  facts combine and result  in a particular set of inner equations.
This set splits into two subsets. 
There are equations which involve in particular derivatives in the direction of the null generators of the null hypersurface. These will be referred to as  {\it transport equations}. 
The remaining  equations only  involve derivatives  in directions which are still tangent to the null hypersurface but  transverse to the null generators.  These will be referred to as {\it inner constraints}.

At most  points of ${\cal N}_p$ none of the frame vectors $e_k$ in the normal  gauge  is tangent to ${\cal N}_p$. To derive from the complete set of  equations subsystems which only contain derivatives in directions tangent to ${\cal N}_p$,  one thus needs to take (point dependent) linear combinations of the equations and the dependent variables.
Whatever  one does to obtain the maximal number of transport equations will  amount in the end  to expressing the equations  in terms of a new frame field on ${\cal N}_p \setminus \{p\}$ which is such  that three of the new frame vectors will be tangent to  
${\cal N}_p \setminus \{p\}$.

We shall describe the procedure and the resulting equations and derive the information which will be needed to construct the desired fields on ${\cal N}_p$ near $p$. 
The following discussion, which works out  some of the considerations at the end of section \ref{null-data} in a systematic way,
makes use of the analysis in \cite{friedrich:pure-rad:1986}, to which we refer for more details.
Let $\{\kappa_a\}_{a = 0,1}$ denote the new spin frame field. If it is chosen such that 
the null vector $\kappa_0  \bar{\kappa}_{0'}$ is tangent to the null generators on 
${\cal N}_p \setminus \{p\}$, the vectors  $\kappa_0  \bar{\kappa}_{1'}$ 
and $\kappa_1  \bar{\kappa}_{0'}$  will be tangent to  ${\cal N}_p \setminus \{p\}$  as well.
Because such a frame field cannot have a  direction independent limit at $p$, particular care has to be taken to construct this frame so near $p$ that the 
resulting equations will still admit a convenient analysis  near $p$. 
It will be required that the frame assumes regular limits at 
the point $p$ if $p$  is approached along the null generators of ${\cal N}_p \setminus \{p\}$.
Let $\kappa_a$ denote such a limit frame at $p$. It can be expanded in terms of the 
normal spin frame $\iota_A$ underlying our earlier analysis in the form 
$\kappa_a = \kappa^A\,_a\,\iota_A$. 
It will be convenient and implies no restriction to assume the spinors $ \kappa^A\,_a$, $a = 0, 1$ or, in other words, the frame transformation matrix  $(\kappa^A\,_a)_{A, a = 0, 1}$ to be normalized such that 
\begin{equation}
\label{kappa-normalization}
\kappa^A\,_a\,\epsilon_{AB}\,\kappa^B\,_b = \epsilon _{ab}, 
\quad \quad 
\kappa^A\,_a\,\tau_{AB'}\,\bar{\kappa}^{B'}\,_{b'} = \tau_{ab'}.
\end{equation}
Here $\tau_{AB'} = \sqrt{2}\,\alpha^0\,_{AA'} 
= \epsilon_A\,^0\,\epsilon_{A'}\,^{0'} + \epsilon_A\,^1\,\epsilon_{A'}\,^{1'}$, 
the quantities $ \epsilon _{ab}$, $\tau_{ab'}$ and  $\alpha^{\mu}\,_{aa'}$ referring to the new frame  take the same numerical values as $\epsilon_{AB}$, $\tau_{AB'}$ and  $\alpha^{\mu}\,_{AA'}$, and the small letter indices are treated in the same way as the large letter indices.

Because we did not specify the null generator along which  the limit was taken, the conditions above characterize in fact a family of frames at $p$. To describe them in detail, denote by $SU(2)$ the Lie group given by the set of complex 
$2 \times 2$-matrices
$(s^a\,_b)_{a,b = 0, 1}$ satisfying the conditions 
\begin{equation}
\label{SU2-def}
s^a\,_c\,\epsilon_{ab}\,s^b\,_d = \epsilon _{bd}, 
\quad \quad 
s^a\,_c\,\tau_{ab'}\,\bar{s}^{b'}\,_{d'} = \tau_{cd'}.
\end{equation}
Any $s \in SU(2)$ can be written in the form
\begin{equation}
\label{SU2-expl}
s = 
\left(\begin{array}{cc}
\alpha\, - \bar{\beta}\\ 
\beta\,\,\, \,\,\,\,\,\bar{\alpha}\\  
\end{array}\right), \quad \quad \alpha, \beta \in \mathbb{C}, \quad |\alpha|^2 + |\beta|^2 = 1,
\end{equation}
and a basis of its Lie-algebra is given by the matrices
\begin{equation}
\label{SU2-gen's}
h = 
\frac{1}{2}\left(\begin{array}{cc}
i  \,\,\,\,\,\,\,\,0\\ 
0\, - i \\  
\end{array}\right), \quad 
u_1 = 
\frac{1}{2}\left(\begin{array}{cc}
0 \quad  i\\ 
i \quad 0\\  
\end{array}\right), \quad 
u_2 = 
\frac{1}{2}\left(\begin{array}{cc}
0\, - 1\\ 
1\,\,\, \,\,\,\,\,0\\  
\end{array}\right). 
\end{equation}
The subgroup of $SU(2)$ consisting of  the matrices 
\begin{equation}
\label{U1-expl}
exp(\phi\,h)= 
\frac{1}{2}\left(\begin{array}{cc}
e^{i\,\frac{\phi}{2}}  \,\,\,\,\,\,\,0\\ 
\,0\,  \,\,\,\,\,\,e^{-i\,\frac{\phi}{2}}  \\  
\end{array}\right), \quad  \quad \phi \in \mathbb{R},
\end{equation}
will be denoted by $U(1)$. Comparing (\ref{kappa-normalization}) with (\ref{SU2-def}) shows 
that  a complete parametrization of the  transformation matrices  $\kappa^A\,_a$ is obtained by setting $\kappa^A\,_a(s) = \delta^A\,_b\,s^b\,_a$ with $s \in SU(2)$. 
The corresponding frame spinors will be denote by $\kappa_a(s)$.

We shall make use of the left invariant vector fields $Z_{u_1}$, $Z_{u_2}$, $Z_h$ generated by 
$u_1$, $u_2$, $h$ and define the operators
\[
Z_+ = - (Z_{u_2} + i\,Z_{u_1}), \quad Z_- = - (Z_{u_2} - i\,Z_{u_1}),
\]
which satisfy the commutation relation  $[Z_+, Z_-] = 2\,i\,Z_h$.
It should be  noted that $SU(2)$ is a real but not a complex analytic Lie group and 
 $Z_{u_1}$, $Z_{u_2}$ must be considered as real vector fields while 
 $Z_+$ and $Z_-$ take values in the complexifications of the tangent spaces of $SU(2)$ and are complex conjugate to each other. If $f$ is a complex-valued function on $SU(2)$ with complex conjugate $\bar{f}$  it holds thus $Z_{\pm}\,\bar{f} = \overline{Z_{\mp}\,f}$.
 In particular, if the $\kappa^A\,_a$ are considered as complex-valued functions on $SU(2)$ as indicated above we get 
 \begin{equation}
 \label{Z-actions}
  Z_+\,\kappa^{A}\,_{0} = 0, \quad \quad
  Z_+\,\kappa^{A}\,_{1} =  \kappa^A\,_0, \quad \quad
  Z_-\,\kappa^{A}\,_{0} = - \kappa^a\,_1, \quad \quad
  Z_-\,\kappa^{A}\,_{1} = 0, 
  \end{equation}
and if $\bar{\kappa}^{A'}\,_{a'}$ is its spinor complex conjugate
we find with the rule above
 \begin{equation}
 \label{cc-Z-actions}
Z_+\,\bar{\kappa}^{A'}\,_{0'} =  - \bar{\kappa}^{A'}\,_{1'},  \quad \quad
Z_+\,\bar{\kappa}^{A'}\,_{1'} =  0,  \quad \quad
Z_-\,\bar{\kappa}^{A'}\,_{0'} =  0,  \quad \quad
Z_-\,\bar{\kappa}^{A'}\,_{1'} =  \bar{\kappa}^{A'}\,_{0'}.
\end{equation}
 Let $c^{\mu}\,_{aa'}(s) = e^{\mu}\,_{AA'}\,\kappa^A\,_a(s)\,\bar{\kappa}^{A'}\,_{a'}(s)$ be the frame field associated with $\kappa_a$ at $p$ and denote by $S^2$ the sphere $ \{x^{\mu} \in T_pM/ x_{\mu}\,x^{\mu} = 0, \sqrt{2}\,x^0 = 1\}$ in the tangent space of $p$. It holds 
\begin{equation}
\label{c00' expl}
x^{\mu}_*(s) \equiv c^{\mu}\,_{00'}(s)
= \alpha^{\mu}_{aa'}\,s^a\,_0\,\bar{s}^{a'}\,_{0'}
\end{equation}
\[
=
\frac{1}{\sqrt{2}}\,( \delta^{\mu}\,_0
+ 2\,Re(\alpha\,\bar{\beta})\,\delta^{\mu}\,_1
+ 2\,Im(\alpha\,\bar{\beta})\, \delta^{\mu}\,_2
+ (|\alpha|^2 - |\beta|^2)\,\delta^{\mu}\,_3),
\]
and  $x^{\mu}_*(s \cdot t) = x^{\mu}_*(s)$ for all $t \in U(1)$.  
The Hopf map 
\[
S^3 \sim SU(2) \ni s \rightarrow 
x^{\mu}_*(s)  \in S^2,
\]
thus associates with the left cosets $s \cdot U(1)$, $s \in SU(2)$ the  null directions 
$x^{\mu}_*(s)$.  It will be  assumed that the frame $\kappa_a(s)$ (resp. $c_{aa'}(s)$) is parallelly propagated along the null geodesic $\tau \rightarrow \tau\, x^{\mu}_*(s)$, 
$\tau \ge 0$,  of ${\cal N}_p$. Because $\iota_A$ (resp. $e_{AA'}$) is a $p$-centered normal frame, it is related to the frame $\kappa_a$ (resp. $c_{aa'}(s)$) along this curve  by the $\tau$-independent transformation $\kappa^A\,_a(s)$ (resp. $\kappa^A\,_a\,\bar{\kappa}^{A'}\,_{a'}(s)$, which corresponds to a rotation in $SO(3, \mathbb{R}) \sim SU(2))/{\{1, -1\}}$ that leaves the direction $e_0$  invariant). While  the null directions $x^{\mu}_*(s)$ are invariant under the action of $U(1)$ the frames $\iota_a$ resp. $c_{aa'}$ are not and our prescription defines in fact  a smooth bundle of frames $\iota_a(\tau, s)$ (resp. $c_{aa'}(\tau, s)$) over 
${\cal N}_p \setminus\{p\}$ with projection $\pi: \iota_{a}(\tau, s) \rightarrow  \tau\, x^{\mu}_*(s)$ 
(resp. $c_{aa'}(\tau, s) \rightarrow  \tau\, x^{\mu}_*(s)$) and structure group $U(1)$ (resp. $U(1)/{\{1, -1\}}$). For simplicity we will concentrate in the following on the bundle of spin frames, the discussion of the bundle of vector frames being very similar.  
The parallel transport of the frames defines lifts of the null geodesics 
$\tau \rightarrow \tau\, x^{\mu}_*(s)$ to this bundle (`horizontal curves'). The   tangent vector field defined by the lifts will be denoted by $\partial_{\tau}$ and $\tau$ will be considered as a coordinate on 
$\tilde{\cal N}_p$. In the limit as $\tau \rightarrow 0$ everything extends smoothly with the limits of the fibers corresponding to the left cosets of $SU(2)$ (in this sense the limit is even preserving the bundle structure). However, while the projection $\pi$   has rank three over points of ${\cal N}_p \setminus\{p\}$, its rank drops to one in the limit to $\pi^{-1}(p)$. 
In the new setting this fact will be reflected by 
the singular behaviour at $\pi^{-1}(p)$ of the frame and the connection coefficients defined below. 
We denote the bundle in the following by $\tilde{\cal N}_p$ and consider it  as a four dimensional smooth manifold with boundary $\pi^{-1}(p)$, the set of frames $c_{aa'}(s)$ at $p$, diffeomorphic to $\mathbb{R}^+_0 \times SU(2)$. 

To discuss the field equations one could choose a local section of the Hopf fibration at $p$  and push it forward with the flow of  $\partial_{\tau}$ to generate a section of $\tilde{\cal N}_p$.
Because the restriction of the projection $\pi$ will then be a $1:1$ map away from $\pi^{-1}(p)$, it will then be obvious how to lift the frame field.
However, apart from a subtlety which will be discussed in the proof of the second part of Proposition \ref{transport-equ} it will in fact be more convenient to  formulate the transport equations as equations  on  $\tilde{\cal N}_p$, as has been done in \cite{friedrich:pure-rad:1986}. 

A suitable lift of the frame field can conveniently be discussed by  
introducing  on   $\tilde{\cal N}_p$ besides $\partial_{\tau}$ vector fields $X_{\pm}$ and $S$. Because the set $\pi^{-1}(p)$
is parametrized by $SU(2)$, the field $Z_{\pm}$ transfer naturally to this set. We set  
\[
X_{\pm} = Z_{\pm}, \quad S = - 2\,i\,Z_h \quad \mbox{on} \quad \pi^{-1}(p),
\] 
and extend these fields to $\tilde{\cal N}_p$ by Lie transport so that 
\[
[\partial_{\tau}, X_{\pm}] = 0, \quad  [\partial_{\tau}, S] = 0.
\]
It follows then that $S$ is tangent to the fibers of $\tilde{\cal N}_p$ and
\[
X_{\pm}\,\tau = 0, \quad [X_+, X_-] = - S.
\]
In fact, the first result follows from $0 = [\partial_{\tau}, X_{\pm}]\,\tau 
= \partial_{\tau}(X_{\pm}\,\tau) -   X_{\pm}\,1= \partial_{\tau}(X_{\pm}\,\tau)$ and the observation 
that $\lim_{\tau \rightarrow 0}X_{\pm}\,\tau = 0$ because the fields $X_{\pm}$ become in this limit tangent to the set $\pi^{-1}(p)$ on which $\tau$ vanishes. The second result follows  because 
it is satisfied in the limit as $\tau \rightarrow 0$ and because the definitions imply that 
$[\partial_{\tau}, [X_+, X_-] + S] = 0$ on $\tilde{\cal N}_p$.
Because the images of the fields $Z_{\pm}$ under the Hopf map are linearly independent, the images of the fields $X_{\pm}$ under the projection 
$\pi$ will be linearly independent for $\tau > 0$ (and sufficiently small that no caustic points will be met).

The scalar fields $\Omega$ and $s$   lift from  ${\cal N}_p$ to $\tilde{\cal N}_p$  by simple  pull-back under the projection map.  The fields $\psi_{ABCD}$ and $\Phi_{ABA'B'}$ are in addition subject to a frame transformation so that they are related to the lifted fields by 
$\psi_{abcd}(\tau, s) 
= \psi_{ABCD}(\tau\,x^{\mu}_*(s))\,\kappa^A\,_a(s) \ldots  \kappa^D\,_d(s)$  and                                           
$\Phi_{aba'b'}(\tau, s) 
= \psi_{ABA'B'}(\tau\,x^{\mu}_*(s))\,\kappa^A\,_a(s) \ldots  \bar{\kappa}^{B'}\,_{b'}(s)$.

\vspace{.2cm}

Only the fields  $c_{aa'} = e_{AA'}\,\kappa^A\,_{a}\,\kappa^{A'}\,_{a'} $ with 
$aa' \neq 11'$  are  tangent to ${\cal N}_p$ at  
the points $\tau\,x^{\mu}_*(s)$ with $\tau > 0$. Lifts of these tangent vector fields on 
${\cal N}_p$ to points of $\tilde{{\cal N}}_p$ are not immediately well defined because the kernel of the projection $\pi$ is one-dimensional. For $\tau > 0$ there exist, however, unique lifts $\tilde{c}_{aa'}$ for $aa' \neq 11'$, i.e. fields satisfying $T\pi(\tilde{c}_{aa'}) = c_{aa'}$, that can be  expanded in  terms of the vector fields $\partial_{\tau}$, $X_+$, $X_-$. Because 
$c_{00'}$ is tangent to the null geodesics of ${\cal N}_p$, it follows then immediately that 
$\tilde{c}_{aa'} = \partial_{\tau}$.  To analyse the precise behaviour of  $\tilde{c}_{aa'}$
as $\tau \rightarrow 0$, we observe that by our earlier discussions 
$e^{\mu}\,_{AA'} = \alpha^{\mu}\,_{AA'} + O(|x|)$ as $x^{\mu} \rightarrow 0$, which gives 
$c^{\mu}\,_{aa'} = \alpha^{\mu}\,_{AA'}\,\kappa^A\,_{a}\,\kappa^{A'}\,_{a'} + O(|\tau|)$  in this limit. For any smooth function $f = f(x^{\mu})$ we find thus with 
$x^{\mu} = \tau\,x^{\mu}_*(s)$ and (\ref{SU2-expl})
\[
f_{,\mu}\,c^{\mu}\,_{aa'} = f_{,\mu}\,\alpha^{\mu}\,_{AA'}\,\kappa^A\,_{a}\,\kappa^{A'}\,_{a'} 
+ O(|\tau|)
\quad \mbox{for} \quad aa' \neq 11'.
\]
To see how this is related to the action of the vector field $X_+$ on the lift of this function 
to $\tilde{\cal N}_p$, we observe that the vector fields $X_{\pm}$ inherit properties of the fields $Z_{\pm}$ such as
(\ref{Z-actions}), (\ref{cc-Z-actions}) and find with (\ref{c00' expl})
\[
X_+f 
= \tau\,f_{,\mu}\,X_+(\alpha^{\mu}\,_{AA'}\,\kappa^A\,_{0}\,\bar{\kappa}^{A'}\,_{0'}) 
=  - \tau\,f_{,\mu}\,\alpha^{\mu}\,_{AA'}\,\kappa^A\,_{0}\,\bar{\kappa}^{A'}\,_{1'}, 
\]
and similarly 
\[
X_-f =  - \tau\,f_{,\mu}\,\alpha^{\mu}\,_{AA'}\,\kappa^A\,_{1}\,\bar{\kappa}^{A'}\,_{0'},
\]
so that we can write
\[
f_{,\mu}\,c^{\mu}\,_{01'} = - \frac{1}{\tau}\,X_+ f + O(|\tau|), \quad \quad 
f_{,\mu}\,c^{\mu}\,_{10'} = - \frac{1}{\tau}\,X_-f + O(|\tau|).
\]
It follows that the lifted fields with $aa' \neq 11'$ must have expansions of the form
\begin{equation}
\label{A-tilde(c)-expansion}
\tilde{c}_{aa'} = \epsilon_a\,^0\,\epsilon_{a'}\,^{0'}\,\partial_{\tau}
- \frac{1}{\tau}\,(\epsilon_a\,^0\,\epsilon_{a'}\,^{1'}\,X_+
+ \epsilon_a\,^1\,\epsilon_{a'}\,^{0'}\,X_-) + c^*_{aa'}, 
\end{equation}
with 
\begin{equation}
\label{star-A-tilde(c)-expansion}
c^*_{aa'} = b_{aa'}\,X_+ + \bar{b}_{aa'}\,X_-
+ r_{aa'}\,\partial_{\tau}, 
\end{equation}
and complex fields $b_{aa'}$ and  $r_{aa'}$ satisfying 
\begin{equation}
\label{B-tilde(c)-expansion}
\bar{r}_{aa'} = r_{aa'}, \quad b_{00'} = 0, \quad r_{00'} = 0, \quad b_{aa'} = O(|\tau|), \quad r_{aa'} = O(|\tau|).
\end{equation}
Because there has not been specified a rule how to  extend the new coordinates and the fields $\tilde{c}_{aa'}$ off  $\tilde{\cal N}_p$, there cannot be given an explicit coordinate expression for the field $\tilde{c}_{11'}$. It should be noted, however, that the field 
$\tilde{c}_{11'}$ is determined on $\tilde{\cal N}_p$ once the fields $\tilde{c}_{aa'}$, 
$aa' \neq 11'$, are known there.

If it is assumed that the relation  $c_{aa'} = e_{AA'}\,\kappa^A\,_{a}\,\kappa^{A'}\,_{a'} $ holds in a full neighbourhood of the point $p$ with an  $x^{\mu}$-dependent transformation matrix $\kappa^A\,_{a}$ and it is used that  $\kappa^a\,_A \equiv \epsilon^{ab}\,\kappa^B\,_b\,\epsilon_{BA}$ satisfies $\kappa^A\,_a\,\kappa^a\,_B = - \epsilon_B\,^A$,  
the well known transformation law
which relates the connection coefficients
$\tilde{\Gamma}_{aa'bc}$ with respect to the frame $c_{aa'}$  to the connection
coefficients $\Gamma_{AA'BC}$ with respect to the frame $e_{AA'}$ is obtained in the form
\[
\tilde{\Gamma}_{aa'bc} =
- \kappa^B\,_b\,\epsilon_{BC}\,\kappa^C\,_{c, \mu}\,
c^{\mu}\,_{aa'} +
\Gamma_{AA'BC}\,\kappa^A\,_{a}\,\kappa^{A'}\,_{a'} \kappa^B\,_{b}\,\kappa^C\,_{c}. 
\]
Under our assumptions the derivatives
$\kappa^C\,_{c, \mu}\,c^{\mu}\,_{aa'}$ are defined on $\tilde{\cal N}_p$ only for $aa' \neq 11'$
so that the formula above can only be used under this restriction.
With  (\ref{Z-actions}) and (\ref{A-tilde(c)-expansion}) it follows then that 
\begin{equation}
\label{A-tilde(Gamma)-expansion}
\tilde{\Gamma}_{aa'bc} = - \frac{1}{\tau}(
\epsilon_a\,^0\,\epsilon_{a'}\,^{1'}\,\epsilon_b\,^1\,\epsilon_{c}\,^{1}\
+ \epsilon_a\,^1\,\epsilon_{a'}\,^{0'}\,\epsilon_b\,^0\,\epsilon_{c}\,^{0})
+ \Gamma_{aa'bc} \quad \mbox{for} \quad aa' \neq 11',
\end{equation}
with a complex-valued field $ \Gamma_{aa'bc}$ that satisfies
\begin{equation}
\label{B-tilde(Gamma)-expansion}
 \Gamma_{00'bc} = 0, \quad \quad  \Gamma_{aa'bc} = O(|\tau|),
\end{equation}
so that
\[
\tilde{\Gamma}_{00'bc} = 0.
\]

In this form the coefficients  lift to $\tilde{{\cal N}}_p$. As discussed in 
\cite{friedrich:pure-rad:1986}, the coefficients $\tilde{\Gamma}_{aa'bc} $ are in fact obtained by contracting the  connection form on the bundle of frames with the frame field $\tilde{c}_{aa'}$.

On  $\tilde{\cal N}_p$  the covariant derivative in the direction of $\tilde{c}_{aa'}$, $aa' \neq 11'$, which will be denoted by $\tilde{\nabla}_{aa'}$, is now given with  
(\ref{A-tilde(c)-expansion}), (\ref{A-tilde(Gamma)-expansion}) by the same rule as known on the base space so that e.g.
\[
\tilde{\nabla}^d\,_{0'}\,\psi_{abcd} = 
\epsilon^{de}\,(\tilde{c}_{e0'}(\psi_{abcd}) - \tilde{\Gamma}_{e0'}\,^f\,_{(a}\,\psi_{bcd)f}).
\]

It will be convenient to introduce  $\Sigma_{AA'} = \nabla_{AA'}\Omega$ as an additional unknown tensor field.
Because no rule has  been specified to  extend the new coordinates and the fields $\tilde{c}_{aa'}$ away from  $\tilde{\cal N}_p$, there cannot be given an explicit coordinate expression for the derivative of $\Omega$ in the direction of $\tilde{c}_{11'}$. Because the field  $\tilde{c}_{11'}$ is determined on $\tilde{\cal N}_p$ once the fields $\tilde{c}_{aa'}$, $aa'  \neq 11'$, are known there, the field 
$\Sigma_{aa'}(\tau, s) = \Sigma_{AA'}\,\kappa^{A}\,_a\,\bar{\kappa}^{B'}\,_{a'}$ can still be discussed as a tensor field on $\tilde{\cal N}_p$.

We are in a position now to obtain the expressions for the transport equations  induced on 
$\tilde{\cal  N}_p$  in the new gauge and to prove the following result.

\begin{proposition}
\label{transport-equ}
In the conformal gauge (\ref{A-conf-gauge}), (\ref{B-conf-gauge})
the transport equations induced on $\tilde{\cal N}_p$  by the conformal field equations and the structural equations uniquely determine the fields $\Omega$, $\Pi$, $\Phi_{aba'b'}$ and
$\psi_{abcd}$ on $\tilde{\cal N}_p$
once  the radiation field 
\begin{equation}
\label{radiation-field-on-tilde-N}
\psi_0(\tau, s) = 
\kappa^A\,_0\,\kappa^B\,_0\,\kappa^C\,_0\,\kappa^D\,_0\,
\psi_{ABCD}|_{x^{\mu} = \tau\,\alpha^{\mu}_{EE'}\,\kappa^E\,_0\,\bar{\kappa}^{E'}\,_{0'}},
\end{equation}
is prescribed there.

\vspace{.1cm}

\noindent
The fields so obtained also satisfy the inner constraint equations on $\tilde{\cal N}_p$.
\end{proposition}

\begin{remark}
A similar result can be obtained in the vacuum case $\Omega = \,1$. The discussion of that case is more complicated than the one below because then the conformal Weyl tensor does 
not necessarily vanish on $\tilde{\cal N}_p$. We do not work out the details here.
\end{remark}

\noindent
{\bf Proof}:
The gauge conditions  (\ref{A-conf-gauge}), (\ref{B-conf-gauge}) read in the present setting 
 \begin{equation}
 \label{lift-A-conf-gauge}
\Omega = \,0, \quad \Sigma_{aa'} = 0, \quad 
  \Pi  =  \Pi _* \equiv 2\,\eta_{00}\, \,\,\,\mbox{on}\,\,\,\,\pi^{-1}(p),
\end{equation}
\begin{equation}
 \label{lift-B-conf-gauge}
 \Phi_{000'0'}= 0, \quad \Lambda = 0 \quad
 \mbox{on} \quad \tilde{\cal N}_{p}. 
\end{equation}
The transport  equations induced by (\ref{Omega-equ}), i.e. the equations which involve the directional derivative $\tilde{c}_{00'}$  imply in particular
\[
\partial_{\tau}\,\Omega = \Sigma_{00'}, \quad \partial_{\tau}\,\Sigma_{00'} = 0,
\]
and thus $\Omega = 0$, $\Sigma_{00'} = 0$ on $\tilde{\cal N}_p$. With this it follows further
\[
\partial_{\tau}\,\Sigma_{01'} = 0, \quad \partial_{\tau}\,\Sigma_{10'} = 0,
\]
whence  $\Sigma_{01'} = 0$,  $\Sigma_{10'} = 0$ on $\tilde{\cal N}_p$.  The transport equations  induced by  (\ref{Omega-equ}), (\ref{s-equ}) then finally  imply
\[
\partial_{\tau}\,\Sigma_{11'} =  \Pi , \quad \partial_{\tau}\, \Pi  = 0,
 \]
and thus  $\Sigma_{11'} = \tau\, \Pi _*$, $ \Pi  =  \Pi _*$ on $\tilde{\cal N}_p$. Collecting results we find 
\begin{equation}
\label{Omega-s-values-on-N}
\Omega = 0,\quad \Sigma_{aa'} = \tau\, \Pi _*\,\epsilon_a\,^1\,\epsilon_{a'}\,^{1'}, \quad
 \Pi  =  \Pi _* \quad \mbox{on} \quad \tilde{\cal N}_p.
\end{equation}

\vspace{.2cm}

The transport equations induced by the torsion free conditions are given by 
\[
0 = \tilde{t}_{bb'\,aa'} = [\tilde{c}_{bb'}, \tilde{c}_{aa'}]
- (\tilde{\Gamma}_{bb'}\,^{ee'}\,_{aa'} - \tilde{\Gamma}_{aa'}\,^{ee'}\,_{bb'})\,\tilde{c}_{ee'},
\]
with $bb' = 00'$ and $aa' \neq 11'$. Inserting here expressions 
(\ref{A-tilde(c)-expansion}), (\ref{A-tilde(Gamma)-expansion}) and setting the factors of 
$\partial_{\tau}$, $X_+$, $X_-$ in the resulting equation separately equal to zero shows that the content of this equation is equivalent to the conditions 

\begin{equation}
\label{b-transp}
\partial_{\tau}b_{aa'} + \frac{1}{\tau}\,b_{aa'} + \frac{1}{\tau}\,\bar{\Gamma}_{aa'0'0'}
= \Gamma_{aa'00}\,b_{10'} + \bar{\Gamma}_{aa'0'0'}\,b_{01'},
\end{equation}

\begin{equation}
\label{r-transp}
\partial_{\tau}r_{aa'} + \frac{1}{\tau}\,r_{aa'} = 
\Gamma_{aa'00}\,r_{10'} + \bar{\Gamma}_{aa'0'0'}\,r_{01'}
- \Gamma_{aa'01} - \bar{\Gamma}_{aa'0'1'},
\end{equation}

\vspace{.2cm}

\noindent
(which are satisfied identically for $aa' = 00'$).

The Ricci identity is given for $cc', dd' \neq 11'$ on $\tilde{\cal N}_p$ by
\[
\tilde{c}_{cc'}(\tilde{\Gamma}_{dd'ab}) - \tilde{c}_{dd'}(\tilde{\Gamma}_{cc'ab}) 
+ \tilde{\Gamma}_{cc'af}\,\tilde{\Gamma}_{dd'}\,^f\,_b  
-  \tilde{\Gamma}_{dd'af}\,\tilde{\Gamma}_{cc'}\,^f\,_b
\]
\[
- ( \tilde{\Gamma}_{cc'}\,^{ff'}\,_{dd'} - \tilde{\Gamma}_{dd'}\,^{ff'}\,_{cc'})\,
\tilde{\Gamma}_{ff'ab}
= \Omega\,\psi_{abcd}\epsilon_{c'd'} + \Phi_{abc'd'}\,\epsilon_{cd}.
\]
With (\ref{Omega-s-values-on-N}),  $\tilde{\Gamma}_{00'ab} = 0$ and $\tilde{c}_{00'} = \partial_{\tau}$ it follows
\[
\partial_{\tau}\tilde{\Gamma}_{10'ab} 
- \tilde{\Gamma}_{10'00}\,\tilde{\Gamma}_{10'ab}
- \bar{\tilde{\Gamma}}_{10'0'0'}\,\tilde{\Gamma}_{01'ab} 
=  \Phi_{ab0'0'},
\]
\[
\partial_{\tau}\tilde{\Gamma}_{01'ab} 
- \tilde{\Gamma}_{01'00}\,\tilde{\Gamma}_{10'ab}
- \bar{\tilde{\Gamma}}_{01'0'0'}\,\tilde{\Gamma}_{01'ab} 
=  0, \quad \quad \,\,\, 
\]
and thus with (\ref{A-tilde(c)-expansion}), (\ref{A-tilde(Gamma)-expansion})

\begin{equation}
\label{A-Gamma-transp-equ}
\partial_{\tau}\Gamma_{10'ab} + \frac{1}{\tau}\left\{\Gamma_{10'ab} -
\Gamma_{10'00}\,\epsilon_a\,^0\,\epsilon_b\,^0     
+ \bar{\Gamma}_{10'0'0'} \,\epsilon_a\,^1\,\epsilon_b\,^1\right\}
\end{equation}
\[
\quad \quad \quad\,\,\, = \Gamma_{10'00}\,\Gamma_{10'ab}
+ \bar{\Gamma}_{10'0'0'}\,\Gamma_{01'ab}
+ \Phi_{ab0'0'},
\]

\begin{equation}
\label{B-Gamma-transp-equ}
\partial_{\tau}\Gamma_{01'ab}  + \frac{1}{\tau}\left\{\Gamma_{01'ab} +
\Gamma_{01'00}\,\epsilon_a\,^0\,\epsilon_b\,^0     
+\bar{\Gamma}_{01'0'0'} \,\epsilon_a\,^1\,\epsilon_b\,^1\right\}
\end{equation}
\[
= \Gamma_{01'00}\,\Gamma_{10'ab} 
+ \bar{\Gamma}_{01'0'0'}\,\Gamma_{01'ab}.
\]

\vspace{.1cm}

\noindent
The transport equations induced by (\ref{Phi-equ}) are
$\tilde{\nabla}_0\,^{c'}\,\Phi_{bcb'c'}  = \psi_{bcd0}\,\Sigma^d\,_{b'}$ or, more explicitly,
\begin{equation}
\label{Phi-transp-equ}
\partial_{\tau}\Phi_{bcb'1'} + \frac{1}{\tau} \left\{ X_+\Phi_{bcb'0'} 
- 2\,\epsilon_{(b}\,^1\,\Phi_{c)0b'0'} + \epsilon_{b'}\,^{0'}\,\Phi_{bc1'0'}
+ \Phi_{bcb'1'}\right\} - c^*_{01'}(\Phi_{bcb'0'})
\end{equation}
\[
= - 2\,\Gamma_{01'}\,^f\,_{(b}\,\Phi_{c)fb'0'} - \bar{\Gamma}_{01'}\,^{f'}\,_{b'}\,\Phi_{bcf'0'} 
- \bar{\Gamma}_{01'}\,^{f'}\,_{0'}\,\Phi_{bcb'f'} - \tau\, \Pi _*\,\psi_{bc00}\,\epsilon_{b'}\,^{1'}, 
\]

\vspace{.1cm}

\noindent
while the transport equations induced by (\ref{psi-equ}) are
$\tilde{\nabla}^d\,_{0'}\,\psi_{abcd} = 0$, or, more explicitly,
\begin{equation}
\label{psi-transp-equ}
\partial_{\tau}\,\psi_{abc1} + \frac{1}{\tau}\left\{X_-\psi_{abc0} +3\,\epsilon_{(a}\,^0\,\psi_{bc)01} + \psi_{abc1}\right\} - c^*_{10'}(\psi_{abc0})
\end{equation}
\[
= - 3\,\Gamma_{10'}\,^f\,_{(a}\,\psi_{bc)f0} - \Gamma_{10'}\,^f\,_0\,\psi_{abcf}.
\]

\vspace{.3cm}

While the initial data at $\tau = 0$ are given for $b_{aa'}$, $r_{aa'}$,
$\Gamma_{aa'bc}$ by 
(\ref{B-tilde(c)-expansion}) and (\ref{B-tilde(Gamma)-expansion}), they still have to be specified for  $\Phi_{aba'b'}$, $\psi_{abcd}$. In principle they can be read off from the formal 
expansions determined earlier but we give a different argument because it sheds some light on the 
content of the equations. It is convenient here to use the `essential components'
$\psi_k = \kappa^A\,_{(a}\,\kappa^B\,_b\,\kappa^C\,_c\,\kappa^D\,_{d)_k}\,\psi_{ABCD}(0))$ 
which are obtained by setting $k$ of the lower indices in brackets equal to $1$ and the remaining ones equal to $0$.
Because the vector fields $X_{\pm}$ approach 
in the limit $\tau \rightarrow 0$ the vector fields $Z_{\pm}$, it follows  with 
(\ref{Z-actions}) and (\ref{radiation-field-on-tilde-N})  
\[
\lim_{\tau \rightarrow 0}X_-\psi_0 = Z_-(
\kappa^A\,_0\,\kappa^B\,_0\,\kappa^C\,_0\,\kappa^D\,_0)\,
\psi_{ABCD}(0) = - 4\,\lim_{\tau \rightarrow 0}\psi_1, 
\]
and, more generally,
\[
\lim_{\tau \rightarrow 0}X_-\psi_k = - (4 - k)\,\lim_{\tau \rightarrow 0}\psi_{k + 1}, 
\quad \quad k = 0, \ldots, 4.
\]
In the notation of (\ref{psi-transp-equ}) this is precisely the relation
\[
\lim_{\tau \rightarrow 0}(X_-\psi_{abc0} +3\,\epsilon_{(a}\,^0\,\psi_{bc)01} + \psi_{abc1}) = 0.
\]
It allows one to determine the initial data $\psi_{abcd}(0)$ from the radiation field
and at the same time ensures that the formally singular term in (\ref{psi-transp-equ}) admits
a  limit as $\tau \rightarrow 0$ along any given  given null generator of 
$\tilde{\cal N}_p$. Similarly one can determine by $X_+$ and $X_-$ 
operations the values of 
$\lim_{\tau \rightarrow 0}\Phi_{aba'b'}$ from 
$\Phi_{000'0'}$ with the result that
\[
\lim_{\tau \rightarrow 0}(X_+\Phi_{bcb'0'} 
- 2\,\epsilon_{(b}\,^1\,\Phi_{c)0b'0'} + \epsilon_{b'}\,^{0'}\,\Phi_{bc1'0'}
+ \Phi_{bcb'1'}) = 0,
\]
so that  the formally singular term  in (\ref{Phi-transp-equ}) admits  a limit along a fixed null generator.
However, because $\Phi_{000'0'} = 0$ on $\tilde{\cal N}_p$ by (\ref{lift-B-conf-gauge}), it follows that
\[
\lim_{\tau \rightarrow 0}\Phi_{aba'b'} = 0.
\]

\vspace{.2cm}

The gauge condition (\ref{lift-B-conf-gauge}) and the vanishing of  the Weyl tensor  on $\tilde{\cal N}_p$  
lead to simplifications. With this (\ref{Phi-transp-equ}) implies
\[
\partial_{\tau}\Phi_{000'1'} + \frac{2}{\tau}\,\Phi_{000'1'}
=  2\,\Gamma_{01'00}\,\Phi_{010'0'} + 2\,\bar{\Gamma}_{01'0'0'}\,\Phi_{000'1'}.
\]
Because $\Phi_{010'0'}$ is by assumption the complex conjugate of $\Phi_{000'1'}$
it follows that 
\begin{equation}
\label{Phi01, Phi10}
\Phi_{000'1'} = 0,\,\,\, \Phi_{010'0'} = 0  \quad \mbox{on} \quad \tilde{\cal N}_p.
\end{equation}

\noindent
Equation (\ref{B-Gamma-transp-equ}) implies the coupled system 
\[
\partial_{\tau}\Gamma_{01'00}  + \frac{2}{\tau}\,\Gamma_{01'00}
= (\Gamma_{10'00} + \bar{\Gamma}_{01'0'0'})\,\Gamma_{01'00}, \quad \quad
\]
\[
\partial_{\tau}\Gamma_{01'01}  + \frac{1}{\tau}\,\Gamma_{01'01} 
= \Gamma_{10'01}\, \Gamma_{01'00} 
+ \bar{\Gamma}_{01'0'0'}\,\Gamma_{01'01},
\]
for  $\Gamma_{01'00}$ and $\Gamma_{01'01}$ whence
\begin{equation}
\label{G0100, G0101}
\Gamma_{01'00} = 0, \quad  \Gamma_{01'01} = 0
\quad \mbox{on} \quad \tilde{\cal N}_p.
\end{equation}

\noindent
With (\ref{lift-B-conf-gauge}), (\ref{Phi01, Phi10}), (\ref{G0100, G0101}) equation  
(\ref{A-Gamma-transp-equ}) implies 
 \[
\partial_{\tau}\Gamma_{10'00} 
= \Gamma_{10'00}\,\Gamma_{10'00},
\]
\[
\partial_{\tau}\Gamma_{10'01} + \frac{1}{\tau}\,\Gamma_{10'01} 
= \Gamma_{10'00}\,\Gamma_{10'01},
\]
from which we conclude that
 \begin{equation}
\label{G1000, G1001}
\Gamma_{10'00} = 0, \quad  \Gamma_{10'01} = 0
\quad \mbox{on} \quad \tilde{\cal N}_p.
\end{equation}
With this the remaining equations of (\ref{B-Gamma-transp-equ}) 
and  (\ref{A-Gamma-transp-equ}) read
\[
\partial_{\tau}\Gamma_{01'11}  + \frac{1}{\tau}\,\Gamma_{01'11} 
= 0,
\]
\begin{equation}
\label{A-Gamma1011-transp-equ}
\partial_{\tau}\Gamma_{10'11} + \frac{1}{\tau}\,\Gamma_{10'11}    
= \Phi_{110'0'},
\end{equation}
which give
\begin{equation}
\label{G0111}
\Gamma_{01'11} = 0, \quad \Gamma_{10'11} = \frac{1}{\tau}\int_0^{\tau} \tau'\,\Phi_{110'0'}\,d\tau'
\quad \mbox{on} \quad \tilde{\cal N}_p.
\end{equation}
With these results it follows  from (\ref{b-transp}), (\ref{r-transp}) that
 \begin{equation}
\label{b01,b10,r01,r10}
b_{aa'} = 0, \quad  r_{aa'} = 0, \quad c^*_{aa'} = 0
\quad \mbox{on} \quad \tilde{\cal N}_p \quad \mbox{for} \quad  aa' \neq 11'.
\end{equation}
With the resulting simplifications equations (\ref{Phi-transp-equ}) read
 \[
 \partial_{\tau}\Phi_{010'1'} + \frac{2}{\tau} \,\Phi_{010'1'} = 0,
\]
\[
 \partial_{\tau}\Phi_{001'1'} + \frac{1}{\tau} \,\Phi_{001'1'}  =  - \tau\, \Pi _*\,\psi_{0000}, 
\]
\[
 \partial_{\tau}\Phi_{011'1'} + \frac{1}{\tau} \left\{ X_+\Phi_{010'1'} 
+ \Phi_{011'1'}\right\}  
= - \tau\, \Pi _*\,\psi_{0001}, 
\]
\[
 \partial_{\tau}\Phi_{110'1'} + \frac{1}{\tau} \left\{ X_+\Phi_{110'0'} 
+ 2\,\Phi_{110'1'}\right\} = 0,
\]
 \[
 \partial_{\tau}\Phi_{111'1'} + \frac{1}{\tau} \left\{ X_+\Phi_{110'1'} 
+ \Phi_{111'1'}\right\}  
= - 2\,\Gamma_{01'11}\,\Phi_{010'1'} - \bar{\Gamma}_{01'1'1'}\,\Phi_{110'0'} 
 - \tau\, \Pi _*\,\psi_{1100}. 
\]
The first three of these equations imply
\begin{equation}
\label{Phi0101,Phi0011,Phi0111}
\Phi_{010'1'} = 0, 
\quad  
\Phi_{001'1'} = - \frac{ \Pi _*}{\tau}\int_0^{\tau} \tau'^2\,\psi_{0000}\,d\tau',
\quad
\Phi_{011'1'} = - \frac{ \Pi _*}{\tau}\int_0^{\tau} \tau'^2\,\psi_{0001}\,d\tau'
\quad \mbox{on} \quad \tilde{\cal N}_p.
\end{equation}
Explicit expressions can also be obtained for the solutions of the remaining equations.
In particular, imposing  the reality conditions, using in the forth equation the expression for $\Phi_{001'1'}$ given by (\ref{Phi0101,Phi0011,Phi0111}),   and observing that 
$X_{\pm}\tau = 0$ gives for $\Phi_{011'1'} $ the 
alternative expression 
\begin{equation}
\label{alt-Phi0111}
\Phi_{011'1'} = \frac{ \Pi _*}{\tau^2}\int_0^{\tau}\left(
\int_0^{\tau'} \tau''^2X_-\psi_{0000}\,d\tau''\right) d\tau'.
\end{equation}
Comparing this with the expression in (\ref{Phi0101,Phi0011,Phi0111}), it is seen that consistency requires  
\[
\partial_{\tau}\psi_{0001}
+ \frac{1}{\tau}\left\{X_-\,\psi_{0000} + 4\,\psi_{0001}\right\} = 0,
\]
which is in fact the first of the equations which follow.

\vspace{.2cm}

\noindent
With the results obtained so far the transport equations (\ref{psi-transp-equ}) 
read 
\begin{equation}
\label{psi0001-transp-equ}
\partial_{\tau}\,\psi_{0001} + \frac{1}{\tau}\left\{X_-\psi_{0000} + 4\,\psi_{0001}\right\} 
= 0,
\quad \quad \quad\quad \quad \quad \,\,\,
 \end{equation}

\begin{equation}
\label{psi0011-transp-equ}
\partial_{\tau}\,\psi_{0011} + \frac{1}{\tau}\left\{X_-\psi_{0010}  + 3\,\psi_{0011}\right\} 
= - \Gamma_{10'11}\,\psi_{0000},
\quad
\end{equation}

\begin{equation}
\label{psi0111-transp-equ}
\partial_{\tau}\,\psi_{0111} + \frac{1}{\tau}\left\{X_-\psi_{0011} 
+ 2\,\psi_{0111}\right\} 
= - 2\,\Gamma_{10'11}\,\psi_{0001},
\end{equation}

\begin{equation}
\label{psi1111-transp-equ}
\partial_{\tau}\,\psi_{1111} + \frac{1}{\tau}\left\{X_-\psi_{0111} + \psi_{1111}\right\} 
= - 3\,\Gamma_{10'11}\,\psi_{0011}.
\end{equation}

\vspace{.2cm}

\noindent
Equation  (\ref{psi0001-transp-equ}) has the regular solution 
\[
\psi_{0001} = - \frac{1}{\tau^4}\,\int_0^{\tau} \tau'^3\,X_-\psi_{0000}\,d\tau'.
\]
With   (\ref{G0111}), (\ref{Phi0101,Phi0011,Phi0111}) one obtains
\begin{equation}
\label{G1011}
\Gamma_{10'11} = - \frac{ \Pi _*}{\tau}\,\int_0^{\tau}\left( \int_0^{\tau'}
\tau''^2\,\bar{\psi}_{0'0'0'0'}\,d\tau''\right) d\tau',
\end{equation}
which allows one to obtain successively integral expressions for the remaining components of $\psi_{abcd}$ on $\tilde{\cal N}_p$. This completes the proof of the first part of the Proposition.

\vspace{.5cm}

Equations  (\ref{Omega-equ}) and  (\ref{s-equ})  imply the inner constraints
\[
0 = \tilde{c}_{01'}(\Sigma_{bb'}) 
- \tilde{\Gamma}_{01'}\,^f\,_b\Sigma_{fb'} 
- \bar{\tilde{\Gamma}}_{01'}\,^{f'}\,_{b'}\Sigma_{bf'} 
+ \Omega\,\Phi_{0b 1' b'} -  \Pi \,\epsilon_{0b}\,\epsilon_{1'b'},
\]
\[
0 = \tilde{c}_{01'}( \Pi ) + \Sigma^{bb'}\,\Phi_{0b 1'b'},
\]
and their complex conjugates. A direct calculation using (\ref{Omega-s-values-on-N}), 
(\ref{G0100, G0101}), (\ref{G1000, G1001}) shows that they are indeed satisfied on $\tilde{\cal N}_p$.

\vspace{.5cm}

There do not arise  inner constraints from (\ref{Phi-equ}), (\ref{psi-equ}).  
Those  which have not been discussed yet contain the operator $\tilde{c}_{11'}$ and thus differentiations in directions transverse to ${\cal N}_p$.

\vspace{.5cm}

Inner constraints are  implied by the torsion-free condition and the Ricci identity.
Formula (\ref{torsion-tensor}) suggests that the torsion free condition should read
on $\tilde{\cal N}_p$
\begin{equation}
\label{wrong-torsion-test}
0 = \left\{[\tilde{c}_{01'}, \tilde{c}_{10'}]
- (\tilde{\Gamma}_{01'}\,^{ee'}\,_{10'} - \tilde{\Gamma}_{10'}\,^{ee'}\,_{01'})\,\tilde{c}_{ee'}\right\}
\end{equation}
There arises, however, a subtlety because the commutator of the fields 
$\tilde{c}_{01'}$ and $\tilde{c}_{10'}$  contributes a component which is tangential to the  
fibers of $\tilde{\cal N}_p$. One way to deal this problem is to follow the torsion-free condition in the form (\ref{torsion-free-test}) and test whether the operator above applied to a function $f$ vanishes if this function is the lift of a scalar function on ${\cal N}_p$, whence constant on the fibres.
For reasons which become clear when we discuss the Ricci identity we prefer a different procedure.
If the operator (\ref{torsion-tensor}) is lifted according to our rules, it should not contain a vertical part and therefore the formula above should be corrected by subtracting the vertical part supplied by the commutator. By (\ref{b01,b10,r01,r10}) the commutator is, however, totally vertical, 
\[
[\tilde{c}_{01'}, \tilde{c}_{10'}] = \frac{1}{\tau^2}\,[X_+, X_-] = - \frac{1}{\tau^2}\,S,
\]
and thus drops out after the correction altogether (as it does if applied to the lift of a scalar function).
A second subtlety arises because the relation above appears to involve the operator $\tilde{c}_{11'}$ which suggests that it is not an inner condition on $\tilde{\cal N}_p$. 
With (\ref{G0100, G0101}), (\ref{G1000, G1001}) and with (\ref{G0111}), which states that 
$\Gamma_{01'11}$ as well as its complex conjugate $\bar{\Gamma}_{10'1'1'}$ vanishes, it follows, however that not only the factor of $\tilde{c}_{11'}$ vanishes but that
$(\tilde{\Gamma}_{01'}\,^{ee'}\,_{10'} - \tilde{\Gamma}_{10'}\,^{ee'}\,_{01'}) = 0$
for arbitrary indices $ee'$.
The inner constraint induced by the torsion free conditions is thus indeed satisfied on $\tilde{\cal N}_p$.

\vspace{.5cm}

The problem arising from the commutator of $\tilde{c}_{01'}$ and $\tilde{c}_{10'}$ also affects the discussion of the inner constraints induced by the Ricci identity. If one calculates the spinor analogue of 
(\ref{gen-comm}), which reads for the components of interest here
\[
(\tilde{\nabla}_{01'} \tilde{\nabla}_{10'} - \tilde{\nabla}_{10'} \tilde{\nabla}_{01'})\lambda^a
= r^a\,_{b01'10'}\,\lambda^b - t_{01'}\,^{ee'}\,_{10'}\,\nabla_{ee'}\lambda^a,
\]
one finds that the second term on the right hand side contains a term of the form
\[
[\tilde{c}_{01'}, \tilde{c}_{10'}](\lambda^a)
- (\tilde{\Gamma}_{01'}\,^{ee'}\,_{10'} - \tilde{\Gamma}_{10'}\,^{ee'}\,_{01'})\,\tilde{c}_{ee'}(\lambda^a).
\]
Performing here  the replacement 
$[\tilde{c}_{01'}, \tilde{c}_{10'}] \rightarrow
[\tilde{c}_{01'}, \tilde{c}_{10'}](\lambda^a) +  \frac{1}{\tau^2}\,S(\lambda^a)$ and then ignoring the torsion term as suggested above, has to be compensated by the replacement 
\[
r^a\,_{b01'10'}\,\lambda^b \rightarrow r^a\,_{b01'10'}\,\lambda^b - \frac{1}{\tau^2}\,S\,\lambda^a,
 \]
of the curvature term. To show that the inner constraint induced by the Ricci identity vanishes, we have to 
take into account  the corrected curvature term.

Under the action of the group $U(1)$ the frame $\kappa_a$ transforms as
$\kappa_a \rightarrow \kappa_b\,(exp(\phi\,h))^b\,_a$ and the components of a spinor field
$\lambda = \lambda^a\,\kappa_a$ transform thus as
$\lambda^a \rightarrow (exp(- \phi\,h))^a\,_b\,\lambda^b$. 
This implies that 
\[
S\,\lambda^a = - 2\,i\,\frac{d}{d\phi}((exp(- \phi\,h)^a\,_b\,\lambda^b)|_{\phi = 0}
= 2\,i\,h^a\,_b\lambda^b,
\]
with $(h^a\,_b)_{a, b = 0, 1}$ denoting the matrix  $h$ in (\ref{SU2-gen's}).
The equation which should be checked thus reads 
\[
0 = \tilde{c}_{01'}(\tilde{\Gamma}_{10'ab}) - \tilde{c}_{10'}(\tilde{\Gamma}_{01'ab}) 
+ \tilde{\Gamma}_{01'af}\,\tilde{\Gamma}_{10'}\,^f\,_b  
-  \tilde{\Gamma}_{10'af}\,\tilde{\Gamma}_{01'}\,^f\,_b
\]
\[
- ( \tilde{\Gamma}_{01'}\,^{ff'}\,_{10'} - \tilde{\Gamma}_{10'}\,^{ff'}\,_{01'})\,
\tilde{\Gamma}_{ff'ab} - \frac{2\,i}{\tau^2}\,h_{ab}
- \Omega\,\psi_{ab01}\epsilon_{1'0'} - \Phi_{ab1'0'}\,\epsilon_{01},
\]
where we set $h_{ab} = h^c\,_b\,\epsilon_{ca}$. In the cases $ab = 00$ and $ab = 01$ a direct calculation using the results obtained above shows that this condition is indeed satisfied on 
$\tilde{\cal N}_p$. The case $ab = 11$ is slightly more difficult. With the given results it readily reduces to the condition 
\[
0 = - \frac{1}{\tau}\,X_+\tilde{\Gamma}_{10'11}- \Phi_{1101'}.
\]
Observing (\ref{G1011}), taking the complex conjugate, and using (\ref{alt-Phi0111})
shows that the condition is indeed satisfied. This proves the second assertion of the Proposition. $\Box$

\subsection{The fields on ${\cal N}_p$ in the normal gauge.}

\vspace{.1cm}

In the first part of this section has been shown that there is associated with the radiation field 
 (\ref{smooth-radiation-field}), which reads in the present notation
\[
\psi_0(\tau, s) = 
\kappa^A\,_0\,\kappa^B\,_0\,\kappa^C\,_0\,\kappa^D\,_0\,
\psi^*_{ABCD}(\tau\,\alpha^{\mu}_{EE'}\,\kappa^E\,_0\,\bar{\kappa}^{E'}\,_{0'}),
\]
a unique set of fields
 \begin{equation} 
 \label{fields-on-tilde-N-p}
 \Omega, \,\,\Sigma_{aa'}, \,\,\Pi,\,\, \Phi_{aba'b'}, \,\,\psi_{abcd}, 
 \quad \mbox{and} \quad \tilde{c}_{aa'}, \,\,\tilde{\Gamma}_{aa'bc}, \,\,\,\,aa' \neq 11', 
\end{equation}
on $\tilde{\cal N}_p$ which satisfy the transport equations and the inner constraints induced by the conformal field equations so that the $0000$ components of $\psi_{abcd}$ coincides with $\psi_0(\tau, s)$. Apart from the explicitly described singular terms of  $\tilde{c}_{aa'}$ and 
$\tilde{\Gamma}_{aa'bc}$ these fields are  
smooth functions of $\tau$ and $s \in SU(2)$. 
On the other hand, it has been shown in sections  \ref{formal expansion} to  \ref{f-and-f-derivatives} that  with  the {\it null data derived from $\psi_0$ at $p$} can be associated fields 
\begin{equation} 
 \label{smooth-fields-near-p}
\hat{\Omega}, \,\,\hat{\Sigma}_{AA'}, \,\,\hat{\Pi}, \,\,
\hat{\Phi}_{ABA'B'}, \,\,\hat{\psi}_{ABCD},  \,\, \hat{e}^{\mu}\,_{AA'},
\,\, \hat{\Gamma}_{AA'BC}, 
\end{equation}
which are defined and smooth  on a neighbourhood of $p$, satisfy at $p$ the conformal field equations at all orders, and which have $\infty$-jets at $p$ which are uniquely determined by this property and the requirement that null data derived from $\psi_0$ at $p$ coincide
with null data  at $p$ derived from 
\[
\hat{\psi}_0(\tau, s) = 
\kappa^A\,_0\,\kappa^B\,_0\,\kappa^C\,_0\,\kappa^D\,_0\,
\hat{\psi}_{ABCD}(\tau\,\alpha^{\mu}_{EE'}\,\kappa^E\,_0\,\bar{\kappa}^{E'}\,_{0'}).
\]
While the Taylor expansions of these functions at $p$ are fixed uniquely, they are fairly arbitrary away from $p$.

To understand the relations between these two sets of fields, we consider the fields 
(\ref{smooth-fields-near-p}) at the points 
$x^{\mu} = \tau\,\alpha^{\mu}_{EE'}\,\kappa^E\,_0\,\bar{\kappa}^{E'}\,_{0'}$ of ${\cal N}_p$ and use the $\tau$-independent frame transformation $\kappa^A\,_a$ employed in section \ref{transport-equ-and-inner-constr} 
to express  the fields
(\ref{smooth-fields-near-p}) in terms of the adapted frame to obtain
on $ \mathbb{R}^+_0 \times SU(2) \sim \tilde{\cal N}_p$  the fields
 \begin{equation} 
 \label{1-second-set-of-fields-on-tilde-N-p}
\hat{\Omega}(\tau, s) 
= \hat{\Omega}(\tau\,\alpha^{\mu}_{EE'}\,\kappa^E\,_0\,\bar{\kappa}^{E'}\,_{0'}),
\quad \quad
\hat{\Pi}(\tau, s) 
= \hat{\Pi}(\tau\,\alpha^{\mu}_{EE'}\,\kappa^E\,_0\,\bar{\kappa}^{E'}\,_{0'}),
\end{equation}
\begin{equation} 
 \label{2-second-set-of-fields-on-tilde-N-p}
\hat{\Sigma}_{aa'}(\tau, s) = \hat{\Sigma}_{AA'}(\tau\,\alpha^{\mu}_{EE'}\,\kappa^E\,_0\,\bar{\kappa}^{E'}\,_{0'})\,\kappa^A\,_a\,\bar{\kappa}^{A'}\,_{a'},
\end{equation}
\begin{equation} 
 \label{3-second-set-of-fields-on-tilde-N-p}
\hat{\Phi}_{aba'b'}(\tau, s) =  \hat{\Phi}_{ABA'B'}( \tau\,\alpha^{\mu}_{EE'}\,\kappa^E\,_0\,
\bar{\kappa}^{E'}\,_{0'})\,
\kappa^A\,_a\,\kappa^B\,_b\,\bar{\kappa}^{A'}\,_{a'}\,\bar{\kappa}^{B'}\,_{b'},
\end{equation}
\begin{equation} 
\label{4-second-set-of-fields-on-tilde-N-p}
\hat{\psi}_{abcd}(\tau, s) =  \hat{\psi}_{ABCD}( \tau\,\alpha^{\mu}_{EE'}\,\kappa^E\,_0\,
\bar{\kappa}^{E'}\,_{0'})\,
\kappa^A\,_a\,\kappa^B\,_b\,\kappa^C\,_c\,\kappa^D\,_d.
\end{equation}
Further, we use
the considerations of section  \ref{transport-equ-and-inner-constr} to derive 
fields $ \hat{c}\,_{aa'}$, $ \hat{\Gamma}_{aa'bc}$, $aa' \neq 11'$, on $ \mathbb{R}^+_0 \times SU(2)$ from  $\hat{e}^{\mu}\,_{AA'}$, $\hat{\Gamma}_{AA'BC}$
which have the meaning and the singularity/regularity structure described in (\ref{A-tilde(c)-expansion}), (\ref{A-tilde(Gamma)-expansion}).

Because the fields (\ref{smooth-fields-near-p}) satisfy the field equations at all orders at $p$
and have only been subject to a coordinate and frame transformation, 
the new fields (\ref{1-second-set-of-fields-on-tilde-N-p}) - (\ref{4-second-set-of-fields-on-tilde-N-p}) must satisfy together with the transformed frame and connection coefficients  
the transport equations and inner constraints  induced on $\tilde{\cal N}_p$ at all orders at 
$p$. The uniqueness property stated in Proposition \ref{transport-equ} thus implies that the Taylor expansion of the fields (\ref{1-second-set-of-fields-on-tilde-N-p})
-  (\ref{4-second-set-of-fields-on-tilde-N-p})
 in terms of $\tau$ at $\tau = 0$ must coincide with the corresponding Taylor expansion of the fields (\ref{fields-on-tilde-N-p}) at $\tau = 0$.

This fact can be expressed in the following way. If the curvature fields given  by  (\ref{fields-on-tilde-N-p}) are transformed  into the normal gauge of section  \ref{gauge-cond} by setting   
 \begin{equation} 
\label{fields-on-tilde-N-p-in-normal-gauge}
\Phi_{ABA'B'}
 = \Phi_{aba'b'}\,\kappa^a\,_A\,\kappa^b\,_B \,\bar{\kappa}^{a'}\,_{A'} \, \bar{\kappa}^{b'}\,_{B'}, \quad    
 \psi_{ABCD} = \psi_{abcd}\,\kappa^a\,_A\,\kappa^b\,_B\,\kappa^c\,_C \,\kappa^d\,_D,
\end{equation}
on ${\cal N}_p$, then
\[
\Phi_{ABA'B'} = \sum_{n = 0}^N \frac{1}{n!}\,\tau^n
\kappa^{E_1}\,_{0}\,\bar{\kappa}^{E'_1}\,_{0'}\,\ldots\,
\kappa^{E_n}\,_{0}\,\bar{\kappa}^{E'_n}\,_{0'}\,
\nabla_{E_1 E'_1}\,\ldots\,\nabla_{E_n E'_n}\Phi_{ABA'B'}(0) + O(|\tau|^{N + 1}),
\]
\[
\psi_{ABCD} = \sum_{n = 0}^N \frac{1}{n!}\,\tau^n
\kappa^{E_1}\,_{0}\,\bar{\kappa}^{E'_1}\,_{0'}\,\ldots\,
\kappa^{E_n}\,_{0}\,\bar{\kappa}^{E'_n}\,_{0'}\,
\nabla_{E_1 E'_1}\,\ldots\,\nabla_{E_n E'_n}\psi_{ABCD}(0) + O(|\tau|^{N + 1}),
\]
for given $N \in \mathbb{N}$,  where the coefficients on the right hand sides are the expansion coefficients associated with the null data derived from $\phi_0$  at $p$  as described in 
sections \ref{null-data} and \ref{formal expansion}. 

One can also transform the 
frame vector fields and the connection coefficients given by (\ref{fields-on-tilde-N-p}) into the normal gauge but more complete information is obtained by using the curvature spinor 
\[
R_{ABCC'DD'} =\Omega\,\psi_{ABCD}\,\epsilon_{C'D'} + \Phi_{ABC'D'}\,\epsilon_{CD},  
\]
supplied on ${\cal N}_p$ by (\ref{fields-on-tilde-N-p-in-normal-gauge})
to integrate the analogues of equations (\ref{Ycontr-torsion-free-cond}) and (\ref{Ycontr-ricci-id}) on ${\cal N}_p$
along the curves $\tau \rightarrow
x^{\mu}(\tau) = \tau\,x^{\mu}_*$, where 
$x^{\mu}_* = \alpha^{\mu}\,_{AA'}\,\kappa^A\,_0\,\bar{\kappa}^{A'}\,_{0'}$ is constant along these curves. 
Let $e^{\mu}\,_{k}$ and $\Gamma_i\,^A\,_B$ denote the frame and connection coefficients which constitute in the normal gauge together with the fields
$\Omega$, $\Pi$  $\Phi_{ABC'D'}$, $\psi_{ABCD}$ supplied by  (\ref{fields-on-tilde-N-p})
initial data  on ${\cal N}_p$  for the conformal vacuum equations and set
$c^{\mu}\,_{k} = e^{\mu}\,_{k} - \delta^{\mu}\,_k$. The restriction of equations  
(\ref{Ycontr-torsion-free-cond}) and (\ref{Ycontr-ricci-id}) to the curves $x^{\mu}(\tau)$ can then be written in the form
\begin{equation}
\label{ODEYcontr-torsion-free-cond}
\tau\,\frac{d}{d\tau}c^{\mu}\,_k
+ c^{\mu}\,_k  
+ c^{\mu}\,_l\,\delta^l_{\nu}\,c^{\nu}\,_k 
+ \Gamma_k\,^i\,_l\,\tau\,X^l_*\,(c^{\mu}\,_i 
+ \delta^{\mu}\,_i) 
= 0,
\end{equation}
\begin{equation}
\label{ODEYcontr-ricci-id}
\tau\,\frac{d}{d\tau}\Gamma_{k}\,^A\,_B
+ \Gamma_{k}\,^A\,_{B}
+ \Gamma_{l}\,^A\,_{B}\,\delta^l_{\mu}\,c^{\mu}\,_{k}
+  \Gamma_{k}\,^{i}\,_{l}\,\tau\,X^l_*\,\Gamma_{i}\,^A\,_B -  R^A\,_{B\,ik}\,\tau\,X^i_* = 0,
\end{equation}
with $X^l_* = \delta^l\,_{\mu}\,x^{\mu}_*$. We are interested here in the solutions which are $C^1$ in 
$\tau$ and satisfy 
\[
c^{\mu}\,_k|_{\tau = 0} = 0, \quad  \Gamma_{k}\,^A\,_{B}|_{\tau = 0} = 0.
\]
If  the left hand sides of the equations are contracted with 
$X^l_*$, the curvature term drops out and one gets for
$c^{\mu} \equiv c^{\mu}\,_k\,X^k_*$ and 
$\Gamma^A\,_B \equiv X^k_*\,\Gamma_{k}\,^A\,_B$
equations which can be written
\[
\tau\,\frac{d}{d\tau}(\tau\,c^{\mu})
+ (\tau^{-1}\,c^{\mu}\,_l)\,\delta^l_{\nu}\,(\tau\,c^{\nu}) 
+ (\tau\,\Gamma^i\,_l)\,X^l_*\,(c^{\mu}\,_i 
+ \delta^{\mu}\,_i) 
= 0,
\]
\[
\tau\,\frac{d}{d\tau}(\tau\,\Gamma^A\,_B)
+ (\tau^{-1}\,\Gamma_{l}\,^A\,_{B})\,\delta^l_{\mu}\,(\tau\,c^{\mu})
+  (\tau\,\Gamma^{i}\,_{l})\,X^l_*\,\Gamma_{i}\,^A\,_B = 0.
\]
Because of the smoothness assumption and the initial conditions we can assume  that 
$\tau^{-1}\,c^{\mu}\,_l$ and $\tau^{-1}\,\Gamma_{l}\,^A\,_{B}$ extend as continuous functions to 
$\tau = 0$. This allows us to conclude 
that 
\[
(e^{\mu}\,_k - \delta^{\mu}\,_k)\,\delta^k\,_{\mu}\,x^{\mu} = 0, \quad
\delta^k\,_{\mu}\,x^{\mu} \,\Gamma_{k}\,^A\,_B = 0 \quad \mbox{along} \quad  x^{\mu}(\tau).
\]
By contracting (\ref{ODEYcontr-torsion-free-cond}) with $x^{\nu}_*\,\eta_{\nu \mu}$
and observing that $\Gamma_k\,^i\,_l\,X^l_* \,\delta^{\mu}\,_i \, x^{\nu}_*\,\eta_{\nu \mu}
= \Gamma_{k\,i\,l}\,\,X^i_*\,X^l_*= 0$,
one gets for $c_k = x^{\nu}_*\,\eta_{\nu \mu}\,c^{\mu}\,_k$ the equation
\[
\frac{d}{d\tau}(\tau\,c_k) 
+ (\tau\,c_l)\,\delta^l_{\nu}\,(\tau^{-1}\,c^{\nu}\,_k)
+ \Gamma_k\,^i\,_l\,X^l_*\,(\tau\,c_i) = 0,
\]
which implies
\[
x^{\nu}\,\eta_{\nu \mu}\,(e^{\mu}\,_k - \delta^{\mu}\,_k) = 0 \quad \mbox{along} \quad  x^{\mu}(\tau).
\]
This shows that the gauge conditions (\ref{x-sigma}), (\ref{sigma-x}), (\ref{spin-nablaX-e=0}) will be 
satisfied on ${\cal N}_p$ by any $C^1$ solution to
(\ref{ODEYcontr-torsion-free-cond}), (\ref{ODEYcontr-ricci-id}).

We know from the explicit calculations above that 
$\Phi_{ADA'D'}\kappa^A\,_a\,\kappa^D\,_0\,\bar{\kappa}^{A'}\,_0
\,\bar{\kappa}^{D'}\,_{d'} = \Phi_{a0\,0'd'} = 0$ on ${\cal N}_p$. This implies that 
\[
R^A\,_{B\,CC'DD'}\,\kappa^B\,_0\,X^{CC'}_* =
\Phi^A\,_{B\,C'D'}\,\kappa^B\,_0\,\bar{\kappa}^{C'}\,_0\,\kappa_D\,_0 = 0 
\quad \mbox{along} \quad  x^{\mu}(\tau).
\]
The contraction of (\ref{ODEYcontr-ricci-id}) with $\kappa^B\,_0$ thus gives
\[
\frac{d}{d\tau}(\tau\,\Gamma_{k}\,^A\,_B\,\kappa^B\,_0)
+ (\tau\,\Gamma_{l}\,^A\,_{B}\,\kappa^B\,_0)\,\delta^l_{\mu}\,(\tau^{-1}\,c^{\mu}\,_{k})
+  \Gamma_{k}\,^{i}\,_{l}\,X^l_*\,(\tau\, \Gamma_{i}\,^A\,_B\,\kappa^B\,_0) = 0,
\]
whence
\[
\Gamma_{k}\,^A\,_B\,\kappa^B\,_0 = 0 \quad \mbox{along} \quad  x^{\mu}(\tau).
\]
Consequently, 
$\Gamma_k\,^{CC'}\,_{DD'}\,X^{DD'}_* 
= \Gamma_k\,^{C}\,_{D}\,\kappa^D\,_0\,\bar{\kappa}^{C'}\,_{0'}
+ \bar{\Gamma}_k\,^{C'}\,_{D'}\,\kappa^C\,_0\,\bar{\kappa}^{D'}\,_{0'} = 0$
along $x^{\mu}(\tau)$ and equation (\ref{ODEYcontr-torsion-free-cond}) reduces to 
\[
\tau\,\frac{d}{d\tau}c^{\mu}\,_k
+ c^{\mu}\,_k  
+ c^{\mu}\,_l\,\delta^l_{\nu}\,c^{\nu}\,_k 
+ \Gamma_k\,^i\,_l\,X^l_*\,\tau\,c^{\mu}\,_i  
= 0
\]
The only $C^1$ solution vanishing at $\tau = 0$  is given by $c^{\mu}\,_k = 0$ and thus
\[
e^{\mu}\,_k = \delta^{\mu}\,_k 
\quad \mbox{whence} \quad g_{\mu \nu} = \eta_{\mu \nu}
\quad \mbox{along} \quad  x^{\mu}(\tau).
\]

\vspace{.5cm}

\noindent
ACKNOWLEDGEMENTS: The author  would like to thank Piotr Chru\'sciel and Tim Paetz for discussions and the Erwin Schr\"odinger Institut for financial support.

}

\end{document}